\documentclass[10pt, journal, letterpaper, twocolumn]{IEEEtran}
\ifCLASSOPTIONcompsoc
\usepackage{cite}
\else
\usepackage{cite}
\fi

\ifCLASSINFOpdf

\else

\fi
\usepackage{adjustbox}
\usepackage{blkarray, bigstrut}
\usepackage{xparse}
\usepackage[table,xcdraw]{xcolor}
 \usepackage{longtable}
 \usepackage{subcaption}
\usepackage{multirow} 
\usepackage{algorithmicx}
\usepackage{algpseudocode}
\usepackage{algorithm}
\usepackage{rotating}
\usepackage{tabularx}
\usepackage{array,multirow}
\usepackage{color}
\usepackage{cuted}
\usepackage{subcaption}
\usepackage{flushend}
\RequirePackage{graphicx}
\RequirePackage{mathptmx}      
\RequirePackage{flushend}
\usepackage[pdfusetitle, pdfauthor={Michael Shell, My institution}]{hyperref}
\usepackage{mathtools}
\usepackage{amssymb}
\usepackage{amsmath}
\usepackage{calc}
 \usepackage{enumitem}
 \usepackage{graphicx,booktabs}
\usepackage{relsize}
\usepackage{stfloats}

\usepackage{amsthm}

\begin{document}

\hyphenation{op-tical net-works semi-conduc-tor}

 \title{Deep Image: A precious image-based deep learning method for online malware detection in IoT Environment}


\author{Meysam~Ghahramani,~Rahim~Taheri,~\IEEEmembership{Member,~IEEE},~Mohammad~Shojafar,\\~\IEEEmembership{Senior~Member,~IEEE},~Reza~Javidan,~Shaohua~Wan,~\IEEEmembership{Senior~Member,~IEEE}
\IEEEcompsocitemizethanks{\protect
\IEEEcompsocthanksitem M. Ghahramani and R. Javidan are with the Computer Engineering and IT Department, Shiraz University of Technology, Shiraz, Iran E-mails:(m.ghahramani@sutech.ac.ir, javidan@sutech.ac.ir)
\protect
\IEEEcompsocthanksitem R. Taheri is with the King’s Communications, Learning and Information Processing (kclip) lab, King’s College London, UK E-mail: rahim.taheri@kcl.ac.uk
\protect
\IEEEcompsocthanksitem M. Shojafar is with the 5G \& 6G Innovation Centre (5GIC \& 6GIC), Institute for Communication Systems, University of Surrey, Guildford, United Kingdom E-mail: m.shojafar@surrey.ac.uk
\protect
\IEEEcompsocthanksitem Sh. Wan is with the School of Information and Safety Engineering, Zhongnan University of Economics and Law, 182 Nanhu Avenue, East Lake High-tech Development Zone, Wuhan, 430073, Hubei, China
 E-mail: shaohua.wan@ieee.org
}
\thanks{Copyright (c) 2022 IEEE. Personal use of this material is permitted. However, permission to use this material for any other purposes must be obtained from the IEEE by sending a request to pubs-permissions@ieee.org}
}

\markboth{Submitted to IEEE Internet of Things Journal, March 2022}
{Ghahramani \MakeLowercase{\textit{et al.}}:XXX}


	\maketitle

\begin{abstract}
The volume of malware and the number of attacks in IoT devices are rising everyday, which encourages security professionals to continually enhance their malware analysis tools. Researchers in the field of cyber security have extensively explored the usage of sophisticated analytics and the efficiency of malware detection. With the introduction of new malware kinds and attack routes, security experts confront considerable challenges in developing efficient malware detection and analysis solutions. In this paper, a different view of malware analysis is considered and the risk level of each sample feature is computed, and based on that the risk level of that sample is calculated. In this way, a criterion is introduced that is used together with accuracy and FPR criteria for malware analysis in IoT environment. In this paper, three malware detection methods based on visualization techniques called the clustering approach, the probabilistic approach, and the deep learning approach are proposed. Then, in addition to the usual machine learning criteria namely accuracy and FPR, a proposed criterion based on the risk of samples has also been used for comparison, with the results showing that the deep learning approach performed better in detecting malware.
\end{abstract}


\section{Introduction}

\IEEEPARstart{I}N today's digital world, with the rise of IoT devices malware development continues at a dizzying rate, despite measures to recognise and combat it. Malware analysis is essential to defend against the sophisticated behaviour of this form of malware. Manual heuristic inspection, on the other hand, is no longer effective or efficient. To address these basic challenges, methods based on behavior-based malware detection mixed with machine learning approaches have been enthusiastically accepted. It applies supervised classifiers to assess their forecasting ability in terms of identifying the most relevant features from the original feature set and balancing a high detection rate with cheap computation cost. While machine learning-based malware detection systems have proved efficient at identifying malware, their shallow learning architecture is still not good enough for complicated malware to be identified.

The exponential growth of malware is owing to the financial rewards connected with malware vulnerabilities such as cryptocurrency coin miners, banking Trojans, and ransomware. Analysts depend largely on machine learning algorithms to discover and classify malware intelligently. Machine learning algorithms will develop models of benign software and malware in order to identify malware in IoT environments. The features that are selected and made available are crucial to the effective utilization of these automation systems. These aspects comprise data acquired from static binary executable files and dynamic malware/software execution behavior.

The majority of deep learning-based windows malware detection algorithms now identify the test input file using static PE file attributes. They also create an image from the whole content of the PE file. Because PE files come in a variety of sizes, a scaling technique is necessary to transform images with different dimensions to a single dimension. The static feature-based detection and classification technique's main drawback is that it is easily evaded by disguised malware.\\
However, most of the present research in malware analysis is based on the assumption that a malware sample is accessible. Frequently, this is not the case. For instance, advanced persistent threat organisations are known to conduct targeted attacks and then self-delete the malware they have planted. 

The methods used in most malware detection systems are focused on increasing accuracy and decreasing FPR. However, in this paper, we show that although these two metrics are very suitable, they are not enough and in some cases even with high accuracy and low FPR, the detection system does not work properly. Therefore, in this paper, a metric is presented that can be used in addition to accuracy and FPR to improve the detection system.As will be demonstrated, in contrast to the research literature, which defines accuracy and FPR in terms of reducing mean squared error, the strategy described in this paper is risk-based. This implies that for each sample, a risk percentage is considered, which is a value between zero and one, and the risk criterion is set and a decision is made using this risk percentage.

\subsection{Motivation and open issues}
In high-risk applications such as malware detection, understanding when a machine learning model is uncertain about its prediction is essential. When an automated malware detection algorithm is uncertain about a sample, the estimated uncertainty can be utilised to flag the sample for investigation by a more computationally demanding algorithm or for human review. Therefore, proposing a method that can predict the result with more certainty is important and it is an open problem that this paper has tried to increase this certainty by introducing a risk criterion.

\subsection{Problem definition}
\begin{figure}
  \centering
  \includegraphics[scale=0.40]{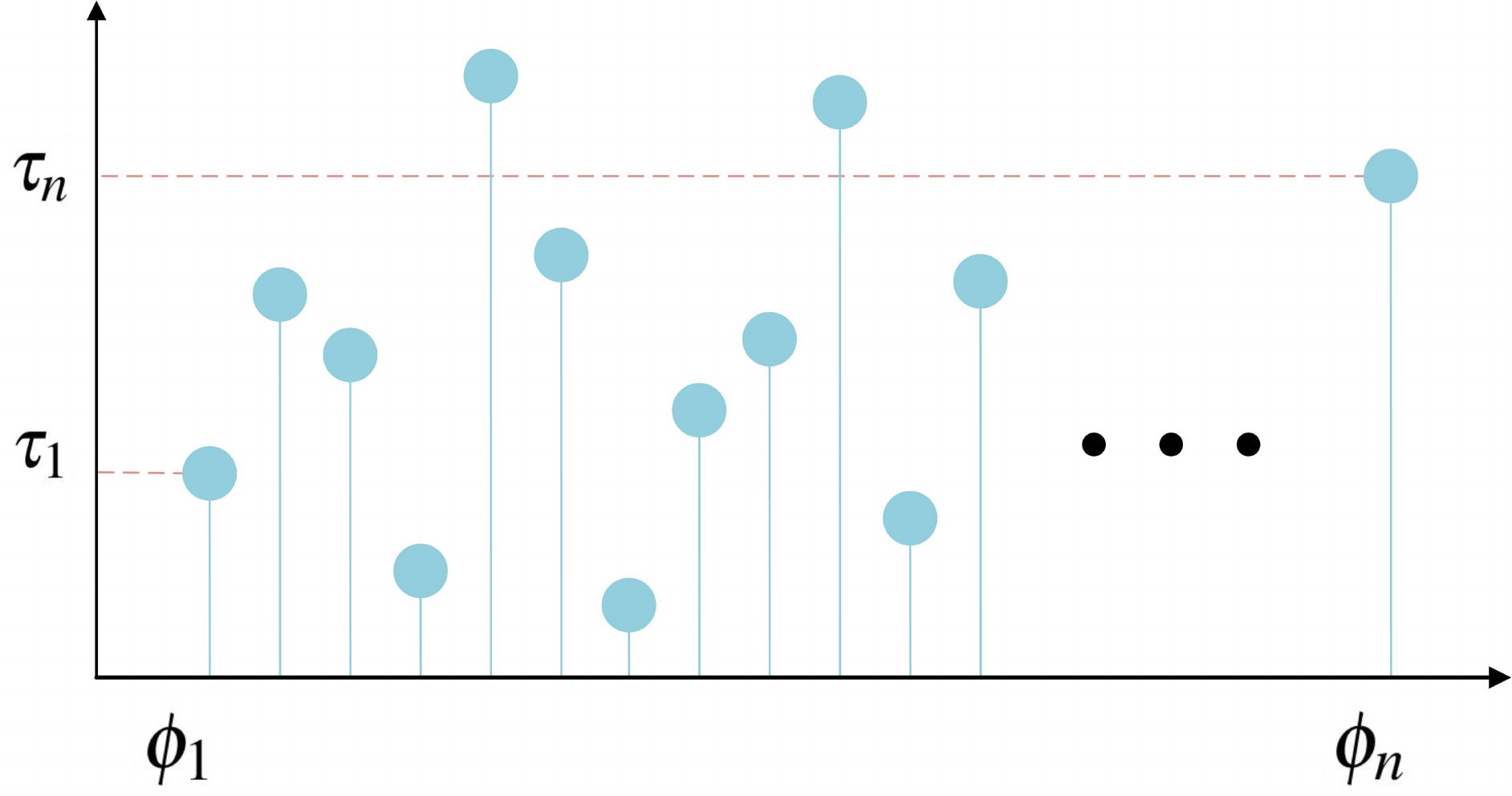}
 \caption{\small 2D form of a time series.}\label{2d}
  \label{cdf}
  \end{figure}

Malware behavior can be demonstrated using time series $\Xi$ shown in eq.~\ref{problem}. In this representation, $0 \leq \xi \leq 1$ shows the maliciousness level. Also, each malware has features $\phi_i$'s, $1\leq i \leq n$, so that the feature $\phi_i$ is available at the time $\tau_i$.
\begin{equation}\label{problem}
\Xi =[\xi, \phi_1 :\tau_1, \phi_2 :\tau_2, \cdots, \phi_n :\tau_n]
\end{equation}
This paper intends to propose a malware detection method to calculate $\xi$ from $[\phi_1 :\tau_1, \cdots, \phi_n :\tau_n]$.
Note that eq.~\ref{problem} is not the only representation for $\Xi$. For example, a two-dimensional diagram with label $\xi$, as in Figure~\ref{2d}, can be used. 
  
The symbols and notation of this paper are summarized in Table~\ref{notations}.
\subsection{Contribution}
To summarise, the following are the most significant contributions made by this paper:
\begin{itemize}[leftmargin=*]
\item We have proposed new visualization-based methods for Windows malware detection.

\item We provide clustering and probabilistic as trivial methods  for predicting the labels of samples that have been converted to images.

\item It is proven that accuracy and false positive rate are not necessarily adequate criteria for estimating the risk weight of malware samples using a criterion that is supplied.

\item A method based on deep learning is proposed that has been converted samples to images samples is presented, which has high performance with both accuracy and FPR criteria and with the introduced risk-based criteria.

\end{itemize}
\subsection{Roadmap}
The remainder of this paper is structured as follows. Section~\ref{relatedWork} provides a brief summary of related studies that have been designed to tackle malware analysis and visualization technique. Section~\ref{proposed} presents proposed architecture,include: clustering, probabilistic and deep learning approaches. A performance analysis of the proposed methods is presented in section~\ref{ExperimentalEvaluation}. Finally, section~\ref{discuss} summarises the main achievements of the paper and gives some directions for future work.

\begin{table}[!t]
\centering
\caption{Notations}\label{notations}
\footnotesize{
\begin{tabular}{cm{5cm}}
Notation&	Description\\
\hline
$\Xi$&	Time series of Malware\\
$0 \leq \xi \leq 1$&  Maliciousness level\\
$\phi_i$,$1\leq i \leq n$&	availability at the time $\tau_i$\\
$\xi$&	label\\
\hline
\end{tabular}}
\end{table}

\section{Related Work}\label{relatedWork}
It has been proposed in the literature a number of ways for malware analysis that are based on machine learning techniques. We will highlight some of the research work that has been done using machine learning techniques to combat Windows malware. Malwares and other malicious software are a collection of instructions designed to do damage to businesses, their processes, networks, and infrastructure. Malware may be either an executable or a non-executable object, and its identification can be accomplished by either static,dynamic or hybrid examination of the infected system.
\subsection{Malware analysis}
Static analysis is the most basic and widely used method for determining the functionality of a system. In this method, the system monitors features of samples and tries to find malwares without running them. It is a technique for extracting features from an executable application, such as APIs,Permissions, hardware components and the Intent of the application~\cite{rizvi2022proud}. These extracted features may be used alone or in combination to detect malware.
Machine learning techniques have been used to find malware in the case of static detection.

Dynamic approaches examine the functionality of an application in running time to determine if it is malicious or not. For example, the paper~\cite{cai2018droidcat} came up with a way to detect Android malware called DroidCat that uses features from method calls and inter-component communication(ICC) to make a multi-class classifier and figure out if an app is malicious or not.

Hybrid approaches combines features of both static and dynamic analysis in a single procedure. Applications that become malware during runtime cannot be detected by static techniques, while malware that conceals its harmful behavior during operating cannot be detected by dynamic methods. Therefore, some authors develop their detection model using a combination of both dynamic and static techniques. They will be able to minimise the downsides while maximising the benefits of both in this manner.

Throughout literature, there are more researches have been done in this area. In addition to ransomware~\cite{fernando2022fesa}, Trojans~\cite{fani2022runtime}, Keyloggers~\cite{singh2021keylogger}, Backdoors~\cite{guo2022backdoor}, Launchers~\cite{cun2020design}, there are other categories of malware. Both signature-based and behavior-based techniques are used for malware detection. The signature-based approach is useful for identifying known threats without producing an excessive number of false alarms~\cite{obaidat2022jadeite}, but it needs regular manual updates of the database containing rules and signatures. According to ~\cite{namanya2020similarity}, a malware detection strategy in the Internet of Things environment based on a similarity hashing algorithm has been presented. The scores of binaries were computed in the proposed approach in order to determine the similarity between malicious PEs.

Hybrid techniques combine aspects of both static and dynamic analysis. Static methods can't find apps that become malware during runtime, and dynamic methods can't find malware that hides its bad behaviour during operation. As a result, some papers use a combination of both dynamic and static approaches to design their detection model. This way, they can minimise the drawbacks and maximise the benefits of both. 

\subsection{Visualization technique}
Image-based malware classification algorithms can overcome code obfuscation or encoding difficulties, comparable to and more advanced than static analysis approaches~\cite{obaidat2022jadeite}. These image-based methods employ static analysis techniques to turn malware samples into image representations~\cite{cho2020mal2d}. Conti et al.~\cite{conti2008visual}conducted the first research on the visualisation of binary data as images. They demonstrated how to convert binary data into visual images, referred known as "byte-plots," in order to increase the capabilities of text-based hex editors. Instead of depending on behaviour or signatures, researchers utilise transformation methods to turn raw malware binaries into pixel-based representations that may be used by machine learning algorithms to classify malware~\cite{cui2019malicious}.

One developing method in machine learning for malware detection is to employ visual features from malware samples in conjunction with powerful machine learning algorithms in computer vision to maximise the benefits of advanced machine learning algorithms~\cite{abdullayeva2019malware,khan2019analysis}. The majority of these methods convert a binary file to an image by reading the binary file's byte sequences as grayscale pixel values.

There was a study done by Yajamanam et al~\cite{yajamanam2018deep} that looked at the GIST-based byte-plot method with a limited set of features. Further testing was done with deep learning and Tensorflow. Results were similar to the original research by Nataraj et al. ~\cite{nataraj2011malware}, but they used a smaller set of 60 features instead of the original 100 features. This is how Le et al.~\cite{le2018deep} used convolutional neural networks to look at binary files from the Microsoft Malware Classification Challenge dataset (Ronen et al.,~\cite{ronen2018microsoft}) that had been turned into byte-plot images. Results from the data show that 98.8\% of the validation data was correct with a processing time of about 20ms per sample.

The researchers in ~\cite{fu2018malware} used local gray-level co-occurrence matrices and global colour moments to extract features from grayscale and colour byte-plots. Classifiers such as Random Forest, Support Vector Machine , and K-Nearest Neighbor were then trained on these features. The studies in this paper used a dataset of 7087 malware samples from 15 various malware families. The Random Forest classifier's accuracy, precision, recall, and f-measure were all more than 97\%.

Unlike the work on byte-plots, there has been relatively little study on how to identify malware using space-filling curves. The authors of ~\cite{baptista2018binary} devised a method for classification malware based on its nature, which makes use of Hilbert curves and a Self-Organizing Incremental Network. Experiments were conducted on 180 samples, with 78 of them proving to be completely innocuous. The malware was classified into these classes: Trojans, ransomware, and unknown. Even though the author claimed that the Hilbert curve was 89\% correct, if you look at just 180 samples, it's not apparent whether the Hilbert curve is an effective method of classifying malware. When working with smaller datasets, overfitting is more likely to occur, which implies the model will not be able to perform effectively with additional samples.

Vu et al.~\cite{vu2020hit4mal} devised a method that combined statistical and syntactic elements, which were then translated into an image format using a variety of layout functions, such as space-filling curves, to create a visual representation of the infection. An experimental dataset of 16,000 excellent and poor images was used to evaluate the technology, which was named hybrid image transformation (HIT). HIT was tested using four distinct layout functions, including two SFC implementations, Hilbert and Hcurve, which were both employed in the testing. According to the authors, their hybrid approach and Hilbert SFC layout function are the most accurate methods for achieving 93 percent accuracy. Throughout this essay, only two categories of objects are considered: those that are good and those that are harmful. It does not introduce malware onto a family's network. When it came to categorization, the only way to determine how well you did was by your accuracy. It would have been possible to utilize other metrics to demonstrate how effectively the method functioned, such as accuracy, recall, and F1- score. It is possible that accuracy alone will not be sufficient to demonstrate how effectively it worked. Aside from that, there was no testing done with a holdout dataset to assess how well the model might generalize to new data.


 
\section{Proposed Method}\label{proposed}
\begin{figure*}
\centering
\includegraphics[scale=0.45]{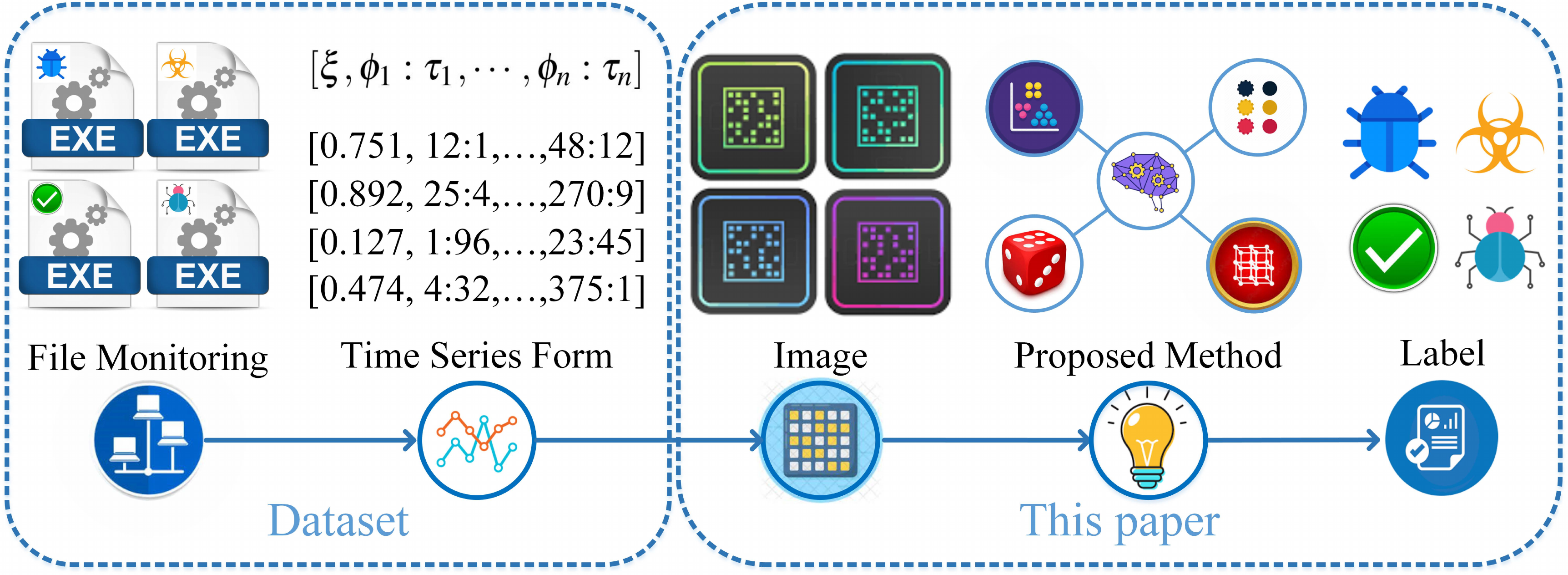}
\caption{\small Architecture of proposed method.}\label{method}
\label{cdf}
\end{figure*}

The method proposed in this paper is dedicated to monitoring target programs during $\tau$ seconds, which is called \textit{monitoring time}. During this time, applications invoke extracted features by the monitoring system as a time series that is accessible using known datasets. The proposed method maps the samples into an image with dimensions $n_\phi \times \tau$ using algorithm~\ref{alg:algorithm}, where $n_\phi$ and $\tau$ represent the number of extracted features and the monitoring time, respectively. After obtaining the images, different methods analyze them, and by introducing a more efficient method, the problems of the previous method are solved. Figure~\ref{method} shows this process.
\subsection{Clustering Approach}
In this section, images are analyzed and their labels are extracted using the clustering technique. To calculate labels, a \textit{clustering image} that contains malware information is needed. This image is a $n_\phi$-by-$\tau \times \ell$ black and white image $CI$ consisting of a number of black or white pixels. This image can be considered as a binary matrix that contains information about $\ell$ clusters. If the element in position $(\phi_i, j)$ is equal to $1$, it means that there is at least \textit{one} malware with label $\xi$, as eq.~\ref{clusteringlabel}, that called feature $\phi_i$ at time $\tau_i=j-\tau \times \lfloor j / \tau \rfloor$.
\begin{equation}\label{clusteringlabel}
\frac{1}{\ell}\times \lfloor\frac{j}{\tau}\rfloor  \leq \xi < \frac{1}{\ell}\times \left(\lfloor\frac{j}{\tau}\rfloor +1 \right)
\end{equation}
Figure~\ref{8clusters} shows a clustering image of the data set used as the train set for proposed clustering approach (CA), which contains more than 80,000 malware for \textit{eight} maliciousness levels. In this figure, the white pixels represent 1, and the black pixels represent 0. To detect malware, it is enough to get a visual representation of it, comparison with the clusters in the clustering image, and report the label of the nearest cluster as a malware label. The proposed clustering approach is summarized in algorithm~\ref{alg:algorithm}.

\begin{algorithm}[t]
\caption{Proposed malware detection algorithm.}
\label{alg:algorithm}
\begin{algorithmic}[1]
\footnotesize
\Statex{\texttt{Required matrices for $\ell$ maliciousness levels:}}
\State{Set a zero $n_\phi$ by $\tau \times \ell$ matrix as $CI$, for clustering image.}
\State{Set a zero $n_\phi$ by $\tau \times \ell$ matrix as $PI$, for Probabilistic image.}
\State{Create a zero $1$ by $\ell$ matrix as $CW$, for the weight of clusters.}
\State{Create a zero $1$ by $\ell$ matrix as $CL$, for the label of clusters.}
\State{Create $\ell$ empty folders named $1,\cdots, \ell$, for storing deep learning images.}
\For{$\Xi \in$ Trainset}
\State{Set the label of $\Xi$ as $\xi$.}
\State{Set the Target Cluster TC as: $TC=\lfloor\ell \times \xi \rfloor +1$.}
\State{$CW_{1 ,TC} \leftarrow CW_{1 ,TC}+1$}
\State{$CL_{1 ,TC} \leftarrow CL_{1 ,TC}+\xi$}
\State{Set a zero $n_\phi$ by $\tau$ matrix as $Z$.}
\For{$(\phi_i:\tau_i) \in \Xi$}
\Statex{\texttt{Time series to image conversion:}}
\State{$Z_{\phi_i , \tau_i} \leftarrow 1$}
\Statex{\texttt{Clustering image computation:}}
\State{$CI_{\phi_i ,(TC-1)\times \tau+ \tau_i} \leftarrow 1$}
\Statex{\texttt{Probabilistic image computation:}}
\State{$PI_{\phi_i ,(TC-1)\times \tau+ \tau_i} \leftarrow PI_{\phi_i ,(TC-1)\times \tau+ \tau_i}+1$}
\EndFor
\Statex{\texttt{Image generation for deep learning detection:}}
\State{Save image form of $Z$ in folder TC.}
\EndFor
\For{$\Xi \in$ Testset}
\State{Set a zero $n_\phi$ by $\tau$ matrix as $Z$.}
\For{$(\phi_i:\tau_i) \in \Xi$}
\State{$Z_{\phi_i , \tau_i} \leftarrow 1$}
\EndFor
\Statex{\texttt{Cluster-based Approach (CA):}}
\State{Compute $i$ that minimizes $\sum_{r=1}^{n_\phi}\sum_{c=1}^{\tau}\vert Z_{r,c}-CI_{r,(i-1)\times \tau+c}\vert$}
\State{Return $\frac{CL_{1,i}}{CW_{1,i}}$ as the estimated maliciousness level $\xi'$.}
\Statex{\texttt{Probability-based Approach (PA):}}
\State{Compute $j$ that minimizes:}
 \Statex{$\quad \quad \sum_{r=1}^{n_\phi}\sum_{c=1}^{\tau}\vert Z_{r,c}-CI_{r,(j-1)\times \tau+c}\times \frac{PI_{r,(j-1)\times \tau+c}}{CW_{1,j}}\vert$}
\State{Return $\frac{CL_{1,j}}{CW_{1,j}}$ as the estimated maliciousness level $\xi'$.}
\Statex{\texttt{Deep learning-based Approach (DA):}}
\State{Find the returned image of Convolution Neural Network (CCN) from folders $1,\cdots,\ell$.}
\State{Return the label of returned image as the estimated maliciousness level $\xi'$.}
\EndFor
\end{algorithmic}
\end{algorithm}

\subsection{Probabilistic Approach}
\begin{figure}
\centering
\includegraphics[scale=0.65]{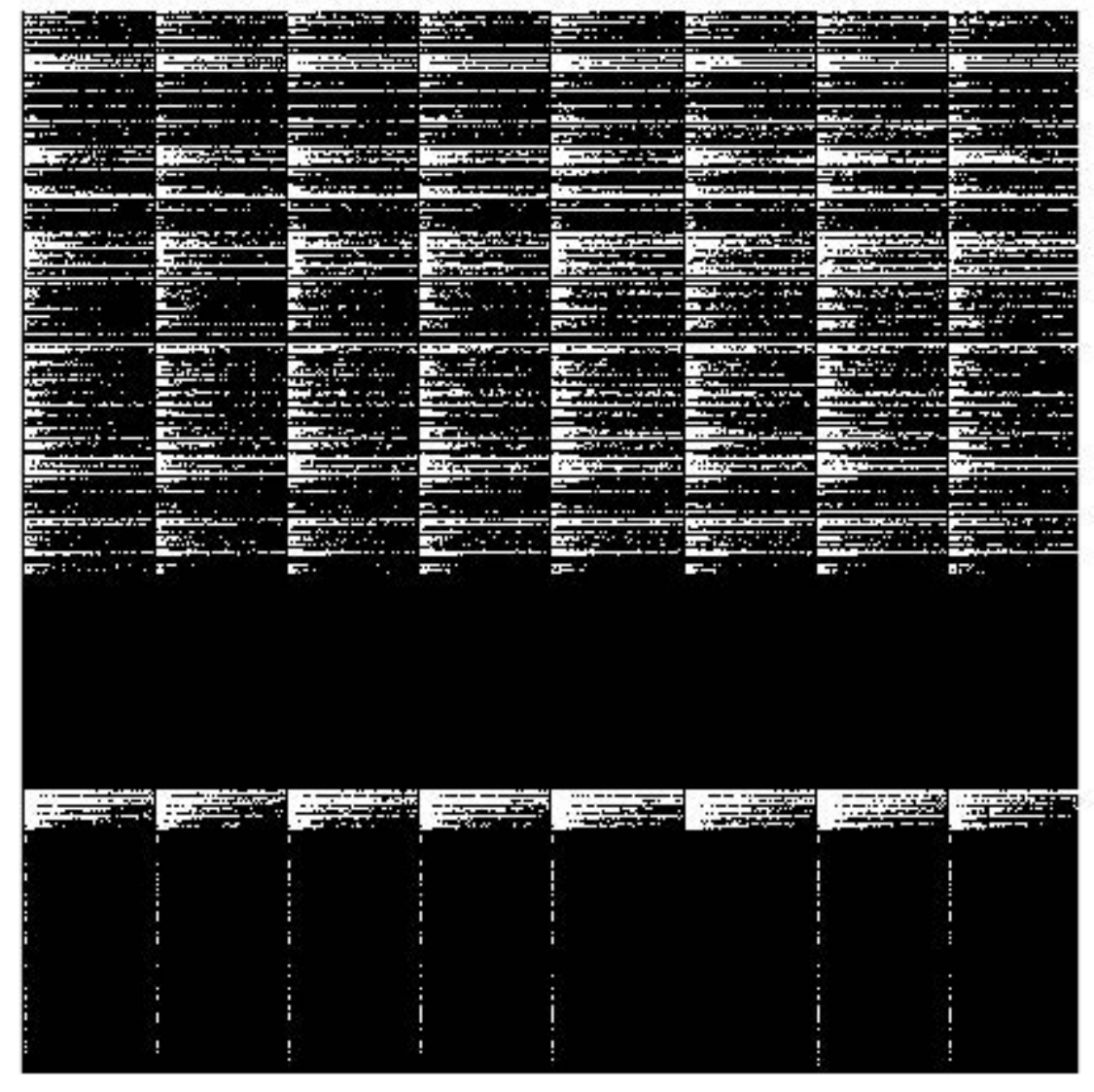}
\caption{\small Clustering image.}
\label{8clusters}
\end{figure}

In the previous section, the clustering approach through clustering image was briefly introduced for malware detection. The clustering image provides suitable information for analysts. For example, at the bottom of Figure~\ref{8clusters} there is a black rectangle, which means that none of the features in this rectangle are called in any of the malware, which can be a useful point in identifying malware behaviours. For this reason, this section proposes another method to improving the performance of clustering image, called the \textit{probabilistic image}. This image is very similar to the clustering image and can be represented by a $n_\phi$-by-$\tau \times \ell$ matrix $PI$ with integer elements. The element $PI_{\phi_i, j}$ on the position $(\phi_i, j)$ indicates that there were $PI_{\phi_i, j}$ malware in the used database, labeled  by $\xi$, which called $\phi_i$ at time $\tau_i=j-\tau \times \lfloor j / \tau \rfloor$. Algorithm~\ref{alg:algorithm} summarizes how to create $PI$, and calculate the malware labels. Similar to the previous method, $\xi$ satisfies eq.~\ref{clusteringlabel}, and the method called Probabilistic Approach (PA).
  
\subsection{Deep Learning Approach}
Although clustering and probabilistic methods can provide valuable information to malware analysts, these methods do not examine or link the different situations. For example, consider a malware that uses feature $\phi_1$ after 1 second of monitoring, and this malware is available in the used database to train the malware detection system. Suppose again that the same malware is present in the test dataset and uses $\phi_1$ after 2 seconds. In this case, the previous two methods verify $CI_{\phi_1 , 2}$ and $PI_{\phi_1 , 2}$ to assess the maliciousness level,  while there is no valuable information in such positions. In other words, previous methods lose their effectiveness against dynamic feature delays. To solve this problem, deep learning-based methods seem to be a good solution. For example, pooling techniques in Convolution Neural Networks (CNN) address adjacent pixels. Therefore, this method converts the samples to images with black and white pixels, stores them according to their maliciousness levels in folders with labels $1, \cdots, \ell$, and examines them by CNN. The details of \textit{three} method are integrated in algorithm~\ref{alg:algorithm}.
  
\subsection{Space \& Time complexity}
The proposed method stores several matrices and images, the sum of their dimensions determines the space complexity. The matrices stored in algorithm~\ref{alg:algorithm} are $CI$, $PI$, $CW$, $CL$ and $Z$, whose dimensions are $n_\phi \times \tau \times \ell$, $n_\phi \times \tau \times \ell$, $\ell$, $\ell$, and $n_\phi \times \tau$, respectively. In addition, the algorithm stores $N_{TR}$ images in different folders, each of which has dimensions $n_\phi \times \tau$. Therefore, the space complexity of algorithm~\ref{alg:algorithm} is $2\times \ell  \times \left(n_\phi \times \tau +1\right)+\left( N_{TR} +1 \right)\times \left( n_\phi \times \tau \right)$.

Similarly, the time complexity of algorithm~\ref{alg:algorithm} can be calculated. The training part of the algorithm consists of \textit{two} loops in lines 6 and 12. So the time complexity of this part is equal to $N_{TE}\times S_\Xi$, where $S_\Xi$ is the size of $\Xi$. At the worst case, $\Xi$ includes $n_\phi \times \tau$ elements. So, $S_\Xi \leq n_\phi \times \tau$. In the test section, there are \textit{two} loops in lines 19 and 21, and the complexity of lines 24, 26, and 28 is assigned to each approach. In line 24, the complexity for calculating $i$ is $n_\phi \times \tau \times \ell$, and in line 26, the complexity of calculating $j$ is the same. Finally, the time complexity of line 28 depends on the complexity of the used deep learning method, which is assumed to be $ T_{CNN}$. As a result, the total space complexity is as eq.~\ref{space}, and eq.~\ref{time} represents the time complexity. 
\begin{equation}\label{space}
O\left( \max \{\ell, N_{TR}\}\times n_\phi \times \tau \right)
\end{equation}
\begin{equation}\label{time}
O\left( N_{TE}\times \left( n_\phi \times \tau \times \ell +T_{CNN} \right)\right)
\end{equation}
The numerical value for these parameters is $\ell =10$, $n_\phi = 482$, $\tau=60$, $N_{TR}=86284$, $N_{TE}=21572$, and $N_{TE}\times T_{CNN}\approx 5318$ seconds per epoch, where $N_{TE}\times T_{CNN}$ is required time for training and testing the dataset using CNN.
\section{Experimental Evaluation}\label{ExperimentalEvaluation}

\subsection{Simulation setup}
\begin{figure}
  \centering
  \includegraphics[scale=0.40]{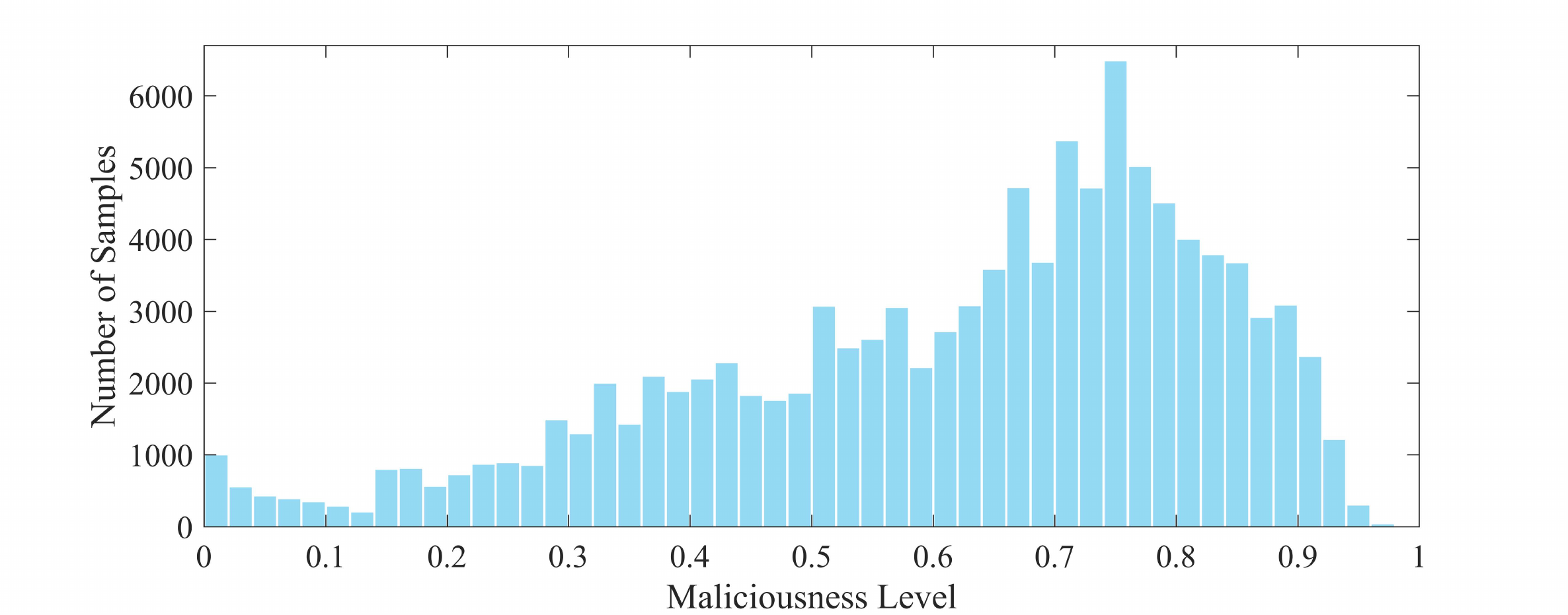}
 \caption{\small Histogram of maliciousness level.}
  \label{histogram}
  \end{figure}

This section compares the proposed method with the state of the art ones by a comprehensive database. The results in this paper are obtained using the database introduced in~\cite{datasetpaper}, which contains more than 100K malware. The malware was obtained by analyzing millions of executable files by 52 antiviruses over a period of 4 years and extracting 486 important features. These features fall into the following four categories: 
\begin{enumerate}
\item Application Programming Interface (API) with 353 features, which is dedicated to accessing the basic functions of the operating system. 
\item Directory with 4 features, which is dedicated to global configuration of operating system.
\item File System with 11 features, which is dedicated to operating system's data organization.
\item Miscellaneous with 118 features, to risk level indication of an executable file.
\end{enumerate}
The used dataset is available on UCI Machine Learning Repository \cite{UCIdataset}. Figure~\ref{histogram} shows the distribution of the maliciousness level $\xi$. As this figure shows, there are more than 6K malware with $0.74<\xi\leq 0.76$. Also, the Cumulative Distribution Function for $\xi$ is depicted in \ref{cdf}, which shows that $100-27.67=72.33\%$ of the analysed malware is very dangerous with $\xi>0.5$. All simulations of this paper are performed in MATLAB R2017a on a laptop with 2.2GHz i7-3632QM CPU and 8GB RAM.  
  
\begin{figure}
  \centering
  \includegraphics[scale=0.26]{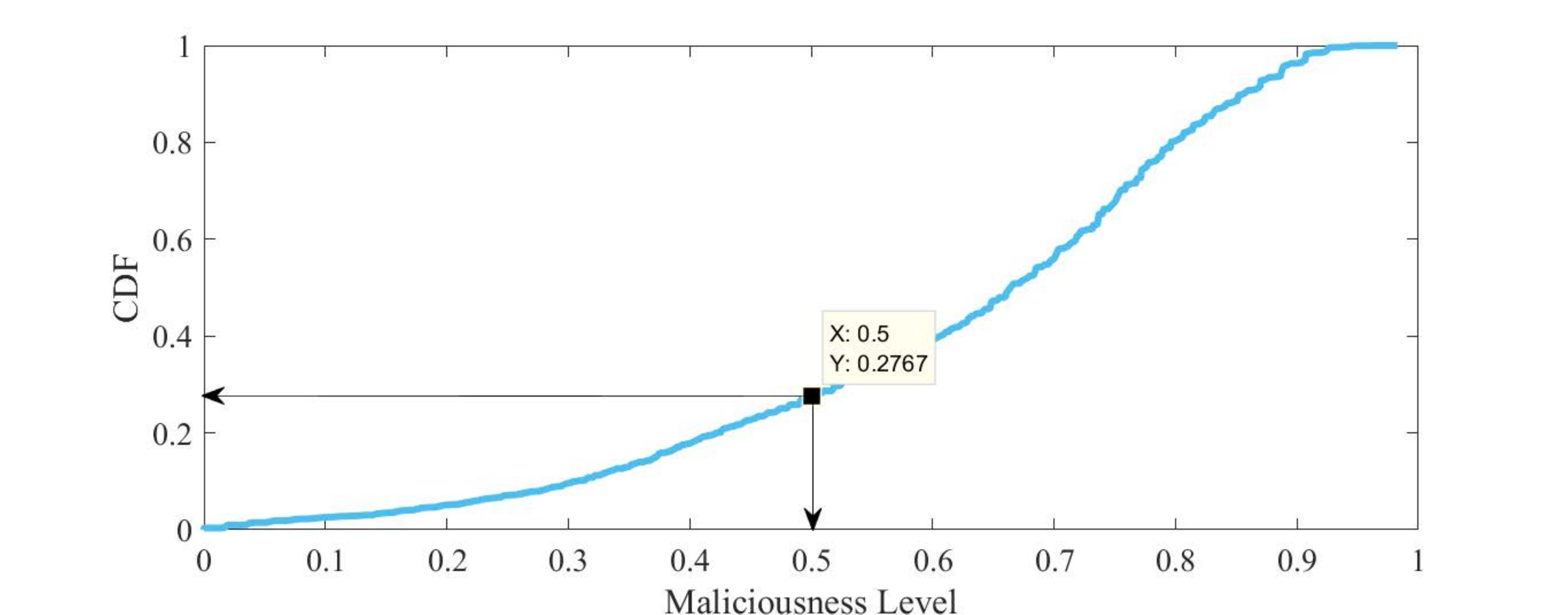}
 \caption{\small CDF of maliciousness level.}
  \label{cdf}
  \end{figure}
\subsection{Comparison of solutions}
In the previous section, \textit{three} approaches to detect maliciousness level of malware were suggested. This section introduces \textit{two} methods for comparison. In 2020, Taheri et al. \cite{similarity} proposed a similarity-based method for detecting malware which is based on Hamming distance. They showed that the static features used in malware are so similar to each other that by examining the similarity of the samples with pre-identified dataset samples, malware can be identified with an accuracy of over 99\%. This method has a better performance compared to fixed size clustering, automatic clustering, neural networks, SVM, etc. For this reason, the method proposed in this paper will be compared with First Nearest Neighbour (FNN) that has the least complexity among the methods in \cite{similarity}. This method is easy to implement, and to calculate the maliciousness level of a test sample, it is enough to calculate its distance from all the members in the trainset. In other words, $ \xi'= \xi_k $, where $1\leq k \leq N_{TR}$ minimizes eq.~\ref{sim}. In this case, $\xi_k$ is the label of $k^{th}$ sample in the trainset with $N_{TR}$ elements. $Z'$ is the image form of the test sample, and $Z^k$ is the the same for $k^{th}$ sample.
\begin{equation}\label{sim}
\sum_{i=1}^{n_\phi}\sum_{j=1}^{\tau}\vert Z'_{i,j} -Z^k_{i,j}\vert
\end{equation}
In 2021, Ritter and Urcid \cite{lattice} published a very interesting study on lattice algebra, which deals with the applications of lattice algebra in image processing, pattern recognition, and artificial intelligence. As a second method to compare with the proposed method, Lattice-based Associated Memory (LAM) is used, which has a stronger mathematical background and much higher efficiency than other methods such as Hopfield Memory. This method integrates the trainset instances into an $n_\phi$-by-$\tau$ image LAM as in~eq.~\ref{lam}, and uses eq.~\ref{latt} to estimate the maliciousness level $\xi'$.
\begin{equation}\label{lam}
LAM_{i,j}= \max \{\xi_k- Z^k_{i,j} \} : 1 \leq k \leq N_{TR}.
\end{equation} 
\begin{equation}\label{latt}
\xi'= \min \{ Z'_{i,j} +LAM_{i,j} \} : 1\leq i \leq n_\phi, 1\leq j \leq \tau.
\end{equation} 
\subsection{Test metrics}
\begin{figure}
\begin{center}
\normalsize{
\setlength\tabcolsep{0.5pt} 
\begin{blockarray}{cccccc}
\begin{block}{c(ccccc)}
$\ell_1$&  $\lambda_{1,1}$ & $\cdots$ & $\lambda_{1,j}$ & $\cdots$ & $\lambda_{1,\ell}$\\
& $\vdots$  & $\ddots$ & $\vdots$&$\ddots$  &$\vdots$ \\
$\ell_i$&  $\lambda_{i,1}$ & $\cdots$ & $\lambda_{i,j}$ &  $\cdots$ &$\lambda_{i,\ell}$\\
& $\vdots$  & $\ddots$ &$\vdots$ & $\ddots$&$\vdots$ \\
$\ell_\ell$&  $\lambda_{\ell,1}$ &  $\cdots$ & $\lambda_{\ell,j}$ &$\cdots$ & $\lambda_{\ell,\ell}$\\
\end{block}
&$\ell_1'$& & $\ell_j'$ & & $\ell_\ell'$ \\
\end{blockarray}
}
\end{center}
\caption{\small Confusion Matrix. $\ell_i$= True label, and $\ell_j'$= Estimated label.}\label{confusion}
\end{figure}
The most well-known evaluation criterion is \textit{mean cumulative absolute error} (MCAE), which is shown in eq.~\ref{mcae}. This criterion shows the average difference between the actual maliciousness level and the estimated one.
\begin{equation}\label{mcae}
\frac{\sum_{i=1}^n \vert \xi_i - \xi'_i \vert}{n}
\end{equation}
In eq.~\ref{mcae}, $n$ represents the size of test set, $\xi_i$ represents the actual maliciousness level of $i^{th}$ test, and $\xi_i'$ represents the estimated one.\\
The next section shows that the MCAE is not sufficient to demonstrate the performance of the proposed method, and it is easy to suggest methods that have a low MCAE but are not suitable for practical purposes and provide unreliable results. For this reason, MCAE and \textit{Confusion Matrix} are used simultaneously in evaluating the proposed method. This matrix displays valuable information about the performance of the proposed method, the structure of which is shown in Figure~\ref{confusion}. In $i^{th}$ row and $j^{th}$ column of confusion matrix there is $\lambda_{i,j}$, which means that in evaluation process there were $\lambda_{i,j}$ samples that had the true label $\ell_i$, while the proposed method estimates label $\ell_j'$. In evaluating the proposed method, the maliciousness levels are divided into 10 parts and if $(i-1)/10\leq \xi_i <i/10$, $\ell_i=i$ is considered, where $1\leq i \leq 10$. The same thing is for $\ell_j'$. 

\begin{figure}
  \centering
  \includegraphics[scale=0.26]{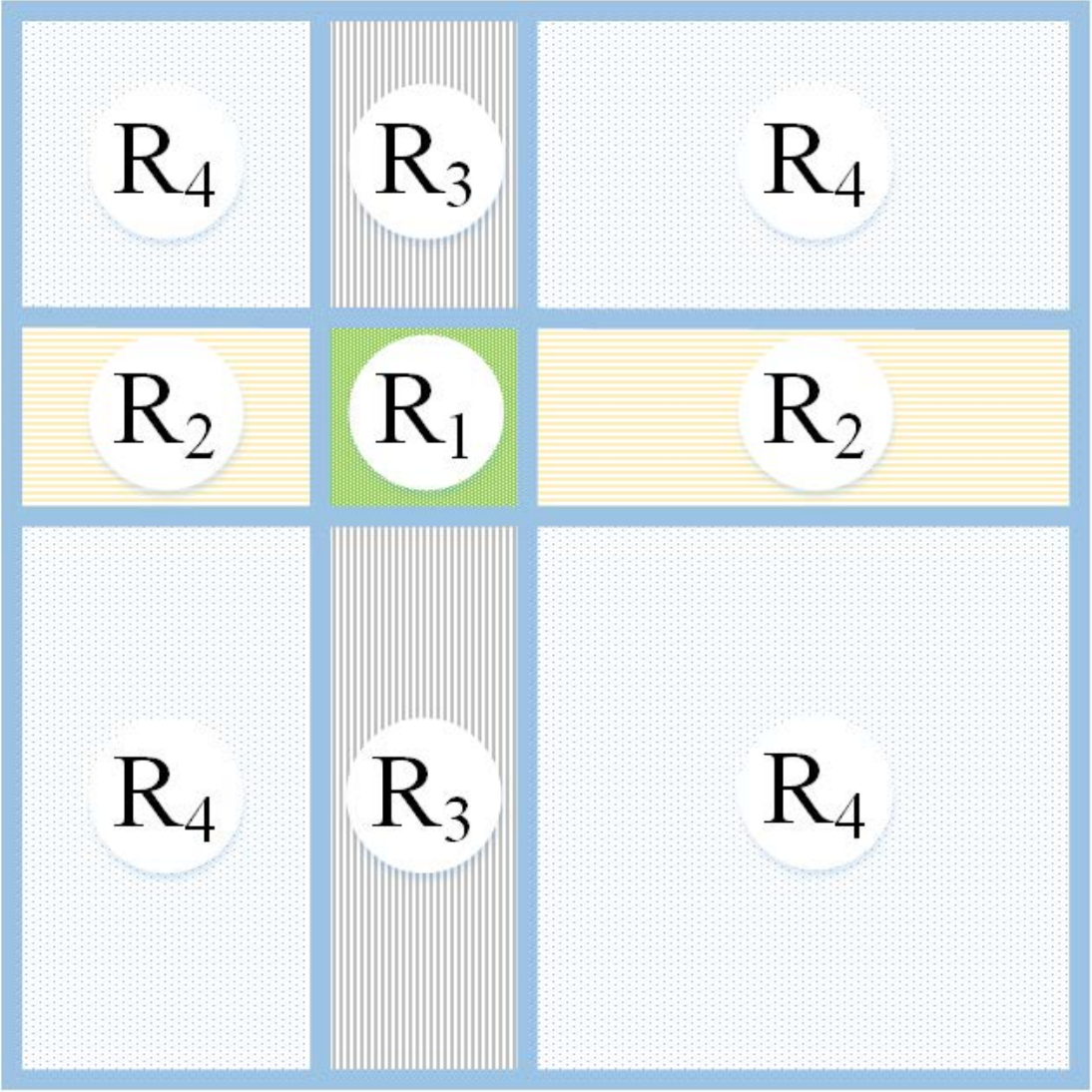}
  \includegraphics[scale=0.31]{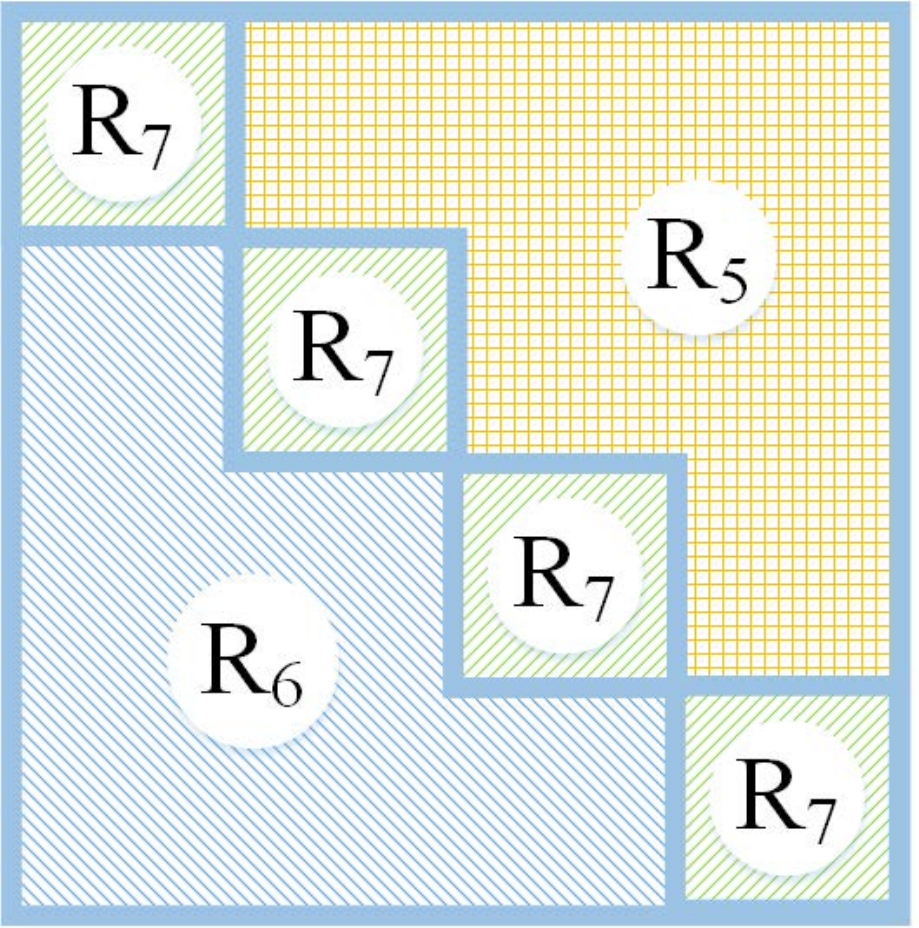}
 \caption{\small Confusion Matrix important regions for test metrics.}
  \label{metrics4}
  \end{figure}
  The performance evaluation criteria by the confusion matrix are based on the \textit{four} numbers shown in Figure~\ref{metrics4}. To calculate these numbers for a target label, the confusion matrix is divided into \textit{four} regions, represented by $R_1$, $R_2$, $R_3$, and $R_4$, which show True Positive, True Negative, False Positive, and False Negative, respectively. Using these values, different test metric can be obtained that show the performance of the proposed method well. Some of these metrics are summarized in Table~\ref{metrictable}.
  \begin{table}[!t]
\centering
\caption{Test metrics for $i^{th}$ label.}\label{metrictable}
\footnotesize{
\begin{tabular}{ccc}
\hline
Metric & Region-based Formula & Confusion Matrix-based Formula\\
\hline
Accuracy & $\frac{R_1+R_4}{R_1+R_2+R_3+R_4}$&$\frac{\lambda_{i,i}+\sum_{r\neq i,r=1}^{\ell}\sum_{c\neq i, c=1}^{\ell}\lambda_{r,c}}{\sum_{r=1}^{\ell}\sum_{c=1}^{\ell}\lambda_{r,c}}$\\
Error Rate & $\frac{R_2+R_3}{R_1+R_2+R_3+R_4}$&$ \frac{\sum_{j\neq i,j=1}^{\ell}\lambda_{i,j}+\lambda_{j,i}}{\sum_{r=1}^{\ell}\sum_{c=1}^{\ell}\lambda_{r,c}}$\\
Precision & $\frac{R_1}{R_1+R_3}$&$ \frac{\lambda_{i,i}}{\lambda_{i,i}+\sum_{c=1}^{\ell}\lambda_{i,c}}$\\
Recall & $\frac{R_1}{R_1+R_4}$&$ \frac{\lambda_{i,i}}{\lambda_{i,i}+\sum_{r\neq i,r=1}^{\ell}\sum_{c\neq i, c=1}^{\ell}\lambda_{r,c}}$\\
F1-Score & $\frac{2R_1}{2R_1+R_3+R_4}$&$ \frac{2\lambda_{i,i}}{2\lambda_{i,i}+\sum_{r\neq i,r=1}^{\ell}\sum_{c=1}^{\ell}\lambda_{r,c}}$\\
\hline
\end{tabular}}
\end{table}
In addition to the metrics introduced in this section, this section introduces two other metrics. A malware detection method is called \textit{conservative} when eq.~\ref{conservative} is established ($R_6 \leq R_5$), otherwise it is called a \textit{loose} detector. In this case, $R_5/R_6$ is called \textit{conservativeness ratio}. Finally, the total accuracy of an algorithm can be expressed by eq.~\ref{acc}.
\begin{equation}\label{conservative}
\frac{R_5}{R_6}=\frac{\sum_{r=1}^{\ell}\sum_{c=r+1}^{\ell}\lambda_{r,c}}{\sum_{r=1}^{\ell}\sum_{c=1}^{r-1}\lambda_{r,c}}\geq 1
\end{equation} 
\begin{equation}\label{acc}
\frac{R_7}{R_5+R_6+R_7}=\frac{\sum_{i=1}^{\ell}\lambda_{i,i}}{\sum_{r=1}^{\ell}\sum_{c=1}^{\ell}\lambda_{r,c}}
\end{equation}
In the next part, the methods introduced in this article are evaluated using the metrics and their pros and cons are examined.
\subsection{Experimental results}
\begin{figure}
  \centering
  \includegraphics[scale=0.25]{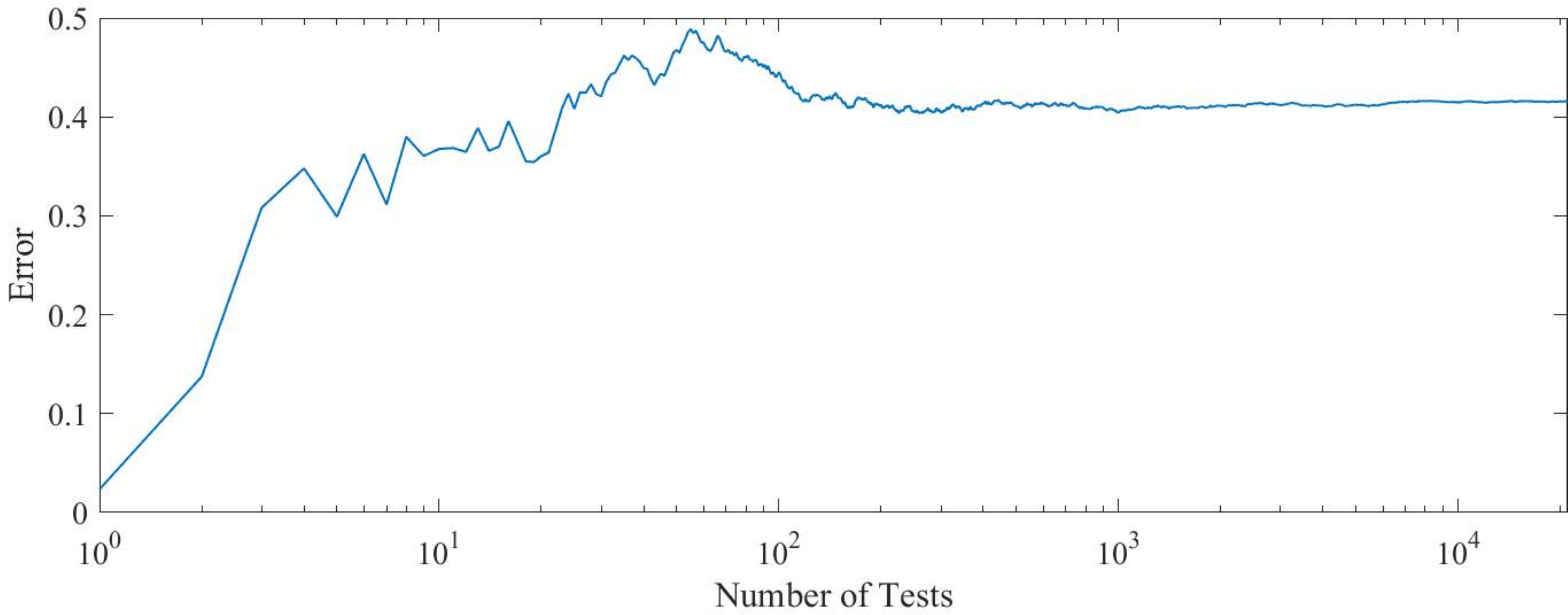}
 \caption{\small MCAE of clustering based approach.}
  \label{clusteringapproach}
  \end{figure}
In this section, the proposed methods are evaluated and the dataset is divided into two parts. The first one is for training and the second, which constitutes 20\% of the data, is used to evaluate. The result of the first experiment is summarized in Figure~\ref{clusteringapproach}, which deals with the MCAE of clustering-based approach. This figure shows that in the first 60 evaluation samples, the average error has an upward trend while it gradually decreases so that after evaluating 400 samples, the average error does not change much. The average error after evaluating 21572 samples is 41.59\%, which is a large error.

  \begin{figure}
  \centering
  \includegraphics[scale=0.38]{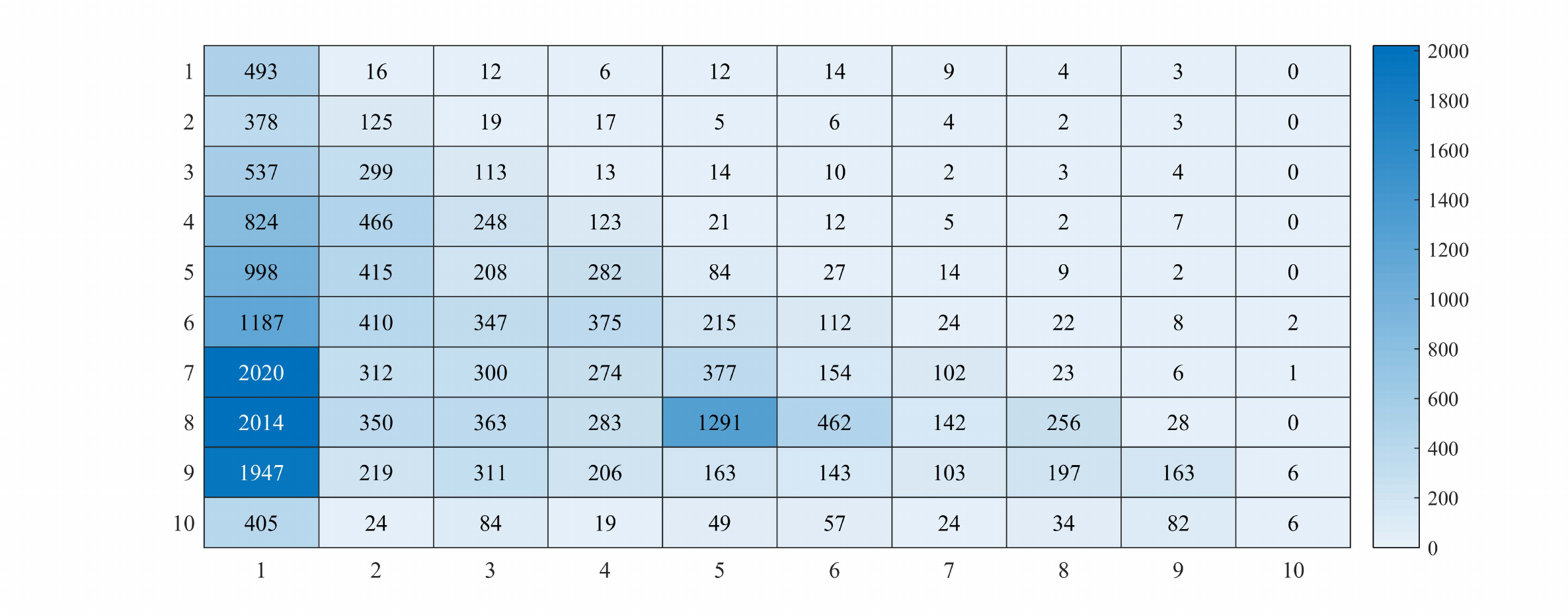}
 \caption{\small Confusion Matrix of clustering based approach.}
  \label{cacm}
  \end{figure}
  Figure~\ref{cacm} shows the Confusion Matrix of this experiment. As this figure shows, $R_6$'s weight is much more than $R_5$, which means that the clustering approach is a loose method. In this case, the probability that the true maliciousness level be less than the estimated one is 5/6=83.33\%, and its total accuracy is only 7.31\%.
  
\begin{figure}
  \centering
  \includegraphics[scale=0.09]{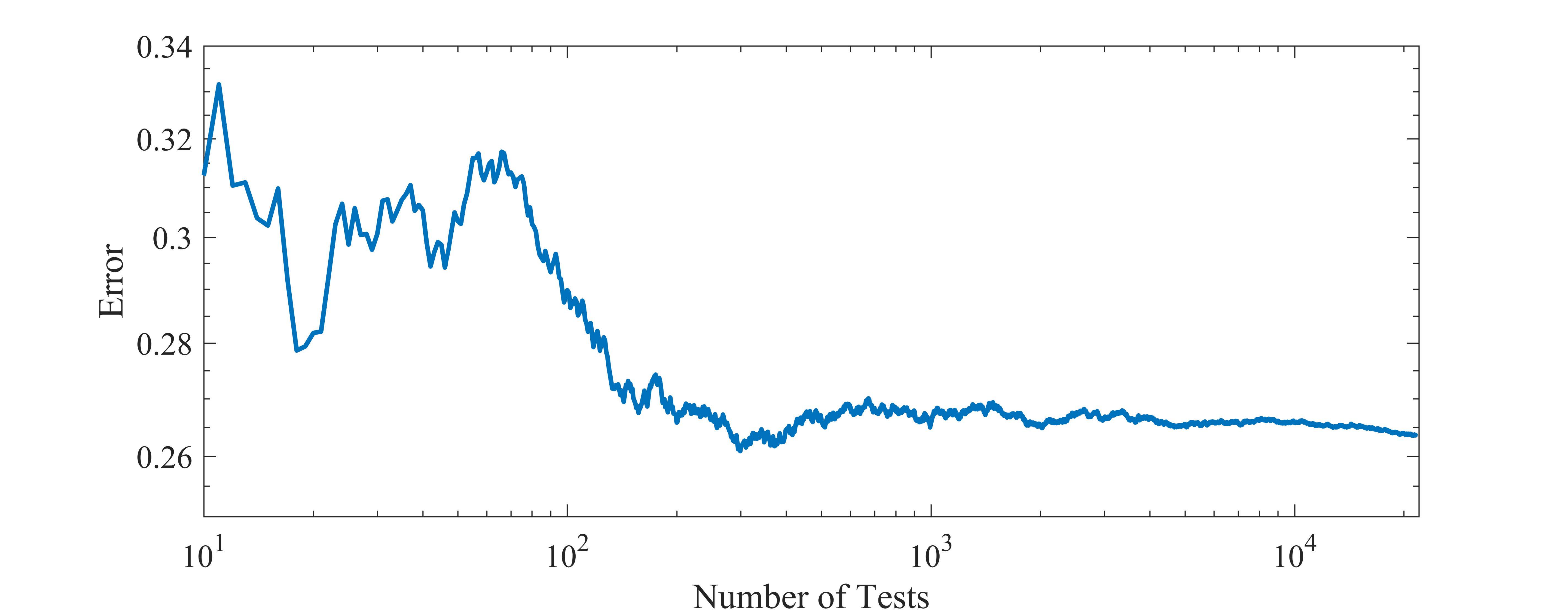}
 \caption{\small MCAE of similarity based approach.}
  \label{simerror}
  \end{figure}
  Figure~\ref{simerror} shows the evaluation results of the similarity-based approach under the same conditions as before, while the error trend is the opposite of the previous approach. The average error trend in this figure is downward and eventually tends to 26.36\%, which improves the previous error by 37\%.
    \begin{figure}
  \centering
  \includegraphics[scale=0.09]{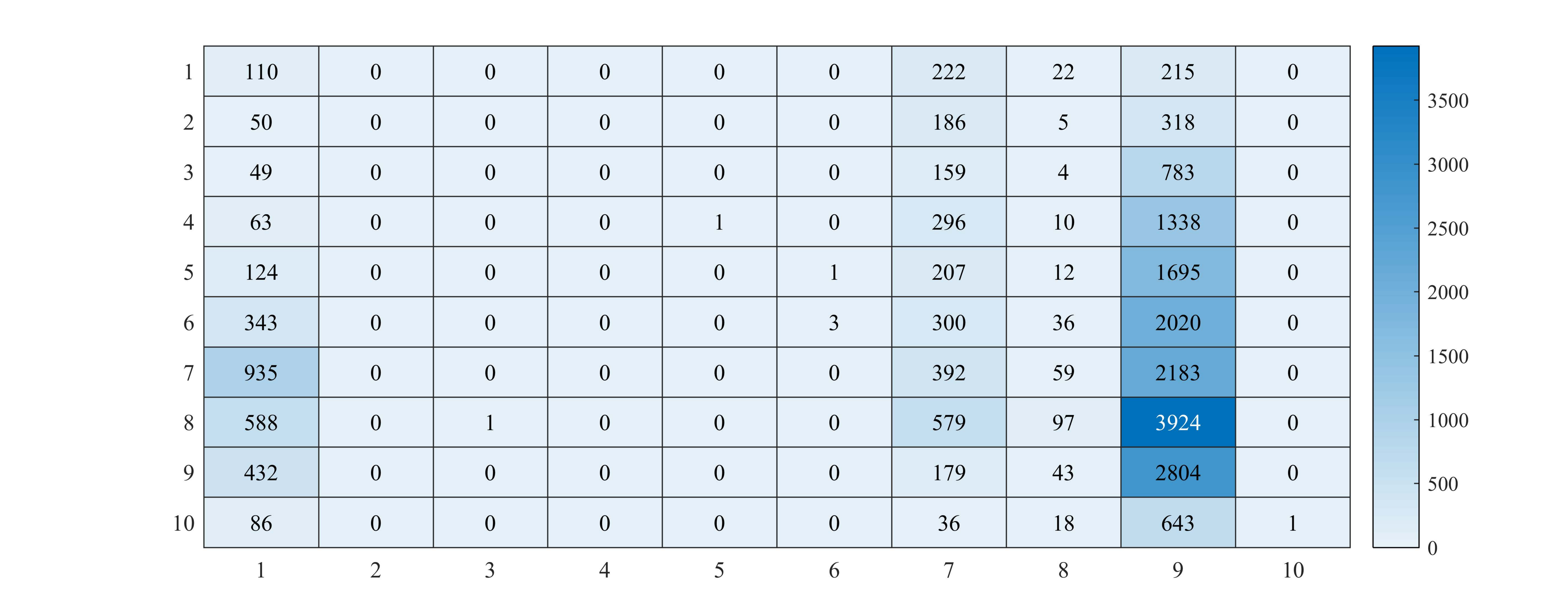}
 \caption{\small Confusion Matrix of similarity based approach.}
  \label{simcm}
  \end{figure}
  Checking the Confusion Matrix gives a similar result. As Figure~\ref{simcm} shows, the weights of $R_5$ is much more than $R_6$, meaning that the similarity approach is a conservative approach and is more likely to identify low-risk malware as high risk one. In other words, $P (\xi '\geq \xi) = 80.67\%$.
  
Examination of the Confusion Matrix of the previous two experiments show a result that cannot be deduced from their MCAE comparisons, as expressed by $R_5$ and $R_6$ regions. Figure~\ref{simcm} shows another result that can be used to improve the MCAE of the proposed methods. As Figure~\ref{simcm} shows, the weights of columns 2 and 4 are zero, which means that no $\xi' \in [0.1, 0.2) \cup [0.3, 0.4)$ is estimated for similarity-based approach. Similarly, the weight of $6^{th}$ column is 4, and the weight of $3^{rd}$, $5^{th}$, and $10^{th}$ columns is one. These results show that the variety of malware with maliciousness level $\xi <0.1$ and $ 0.6 \leq \xi < 0.9 $ is so great that there is always a sample in them that bears the closest similarity to the samples being evaluated. The total accuracy for similarity-based approach is 15.79\%, which is so better in comparison to clustering one.

  \begin{figure}
  \centering
  \includegraphics[scale=0.37]{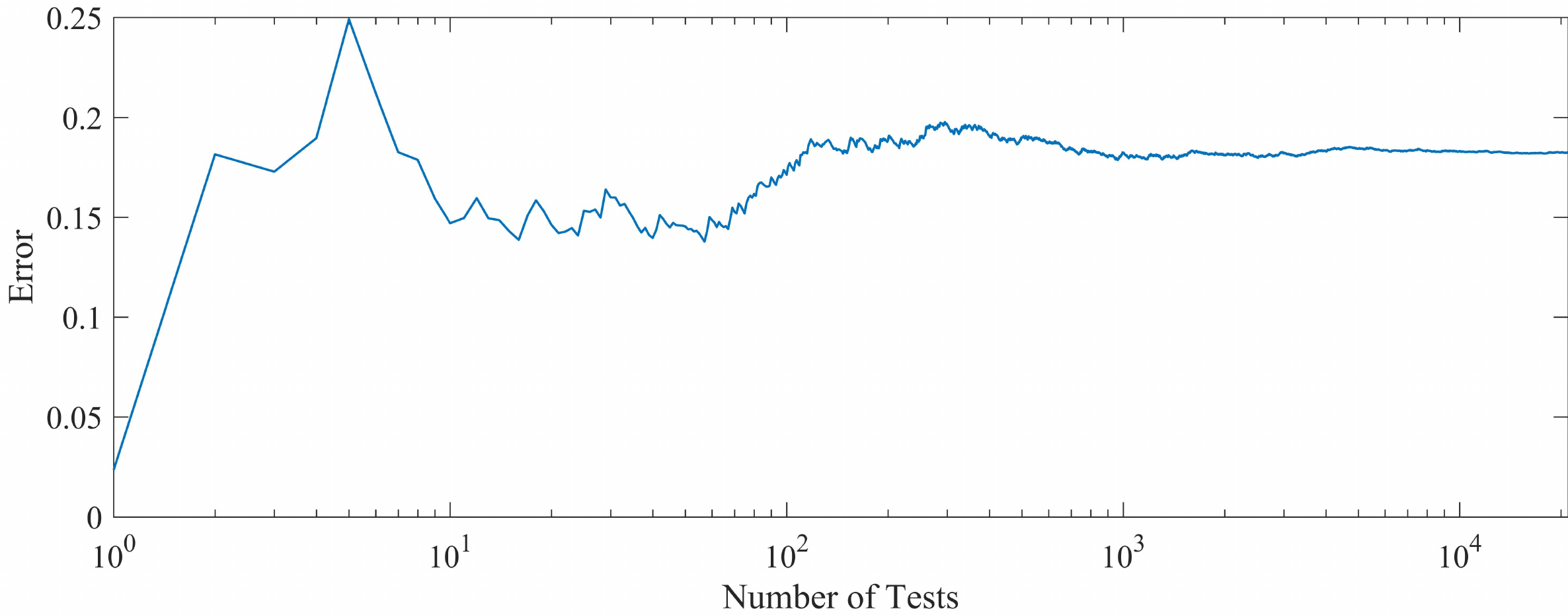}
 \caption{\small MCAE of probability based approach.}
  \label{pba}
  \end{figure}
    \begin{figure}
  \centering
  \includegraphics[scale=0.37]{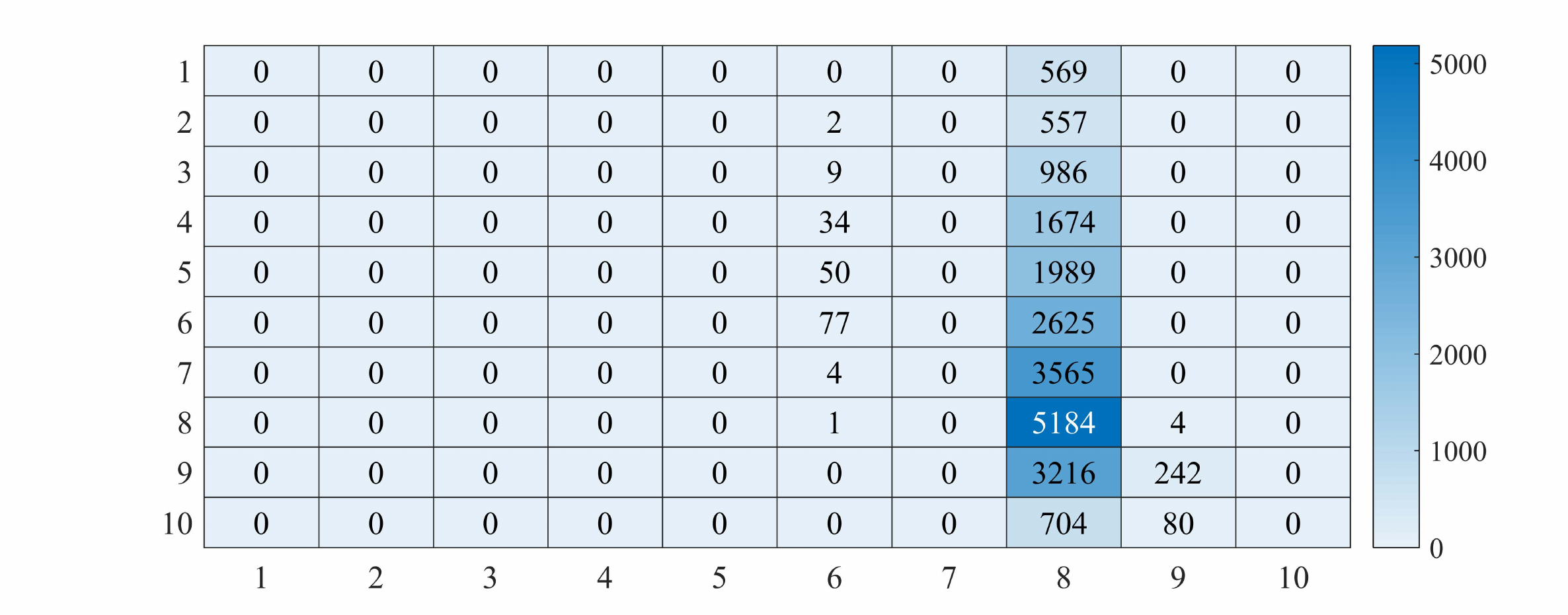}
 \caption{\small Confusion Matrix of probability based approach.}
  \label{pba2}
  \end{figure}
  Figures \ref{pba} and \ref{pba2} show the results of the probabilistic approach evaluation. The general trend of MCAE in this approach is similar to the clustering method, while its Confusion Matrix behavior follows the similarity-based method. The average error of this method fluctuates per 1000 evaluated samples, later it does not change significantly, and eventually converges to 17.6\%, which has less error compared to the previous two approaches. As Figure~\ref{pba2} shows, the Confusion Matrix of this approach has a large number of zero columns, which was also observed in the similarity-based one. In this figure, there are only \textit{three} non-zero columns $C_5$, $C_8$, and $C_9$, where $C_8$ has more weight. This figure shows that with 97.67\% probability, the probabilistic approach identifies all evaluated malware as a malware with maliciousness level $0.7 \leq \xi' <0.8$. The total accuracy of this method is 25.51\%, which has a much better performance compared to the previous two approaches.
  
The number of non-zero columns in the lattice-based Confusion Matrix  reaches its peak so that there is only \textit{one} non-zero column in this matrix, which means that the output of this method with is always a constant maliciousness level $c$. By analyzing eq.~\ref{lam} and eq.~\ref{latt}, the constant output of this method can be calculated. In the used database, $Z'_{i,j}, Z^k_{i,j} \in \{0,1\}$ and $0 \leq \xi <1$. Assume that in eq.~\ref{latt}, $LAM_{i,j}$ is a fixed number $c$. In this case, since  $Z'_{i,j} \in \{0,1\}$ and $LAM_{i,j}=c$, so $\xi'=\min \{0+c, 1+c\}=c$ means that the output of this method is always equal to $c$. In evaluating lattice-based approach, the method returns $c=0.9828$.  Now the question is:
 \textit{If a particular method, such as in lattice-based approach, always provides a fixed output $c$, is this output always the best possible output that keeps the error to minimum?}
 
  \begin{figure}
  \centering
  \includegraphics[scale=0.37]{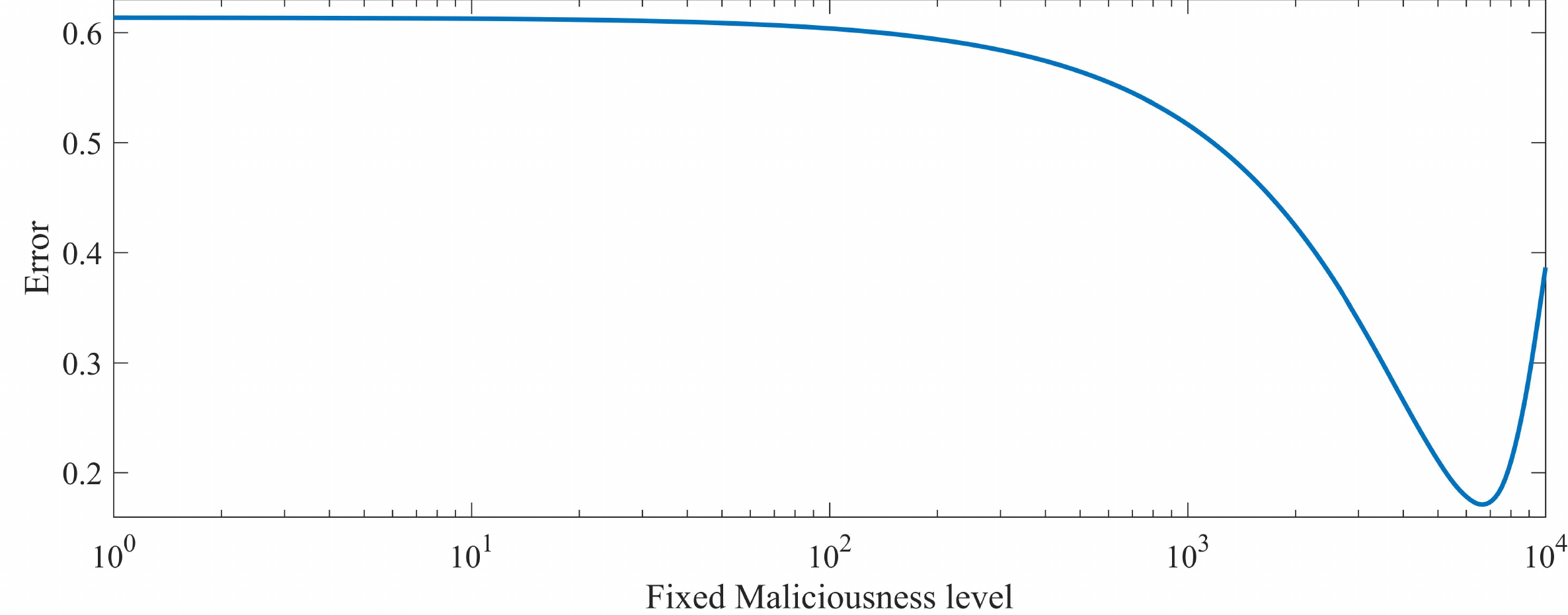}
 \caption{\small MCAE of single level based approach.}
  \label{constant}
  \end{figure}
  Figure~\ref{constant} shows the MCAE for 10000 fixed maliciousness levels. As this figure shows, for $c = 0$, the lattice-based method is a loose approach that identifies all dangerous malware as harmless, the MCAE is over 60\%, which is very high. As $c$ increases, the mean error decreases so that at $0.6613 \leq c \leq 0.6697$ the mean error reaches the lowest possible value of 17.1\%. As shown in Figure~\ref{cdf}, 72.33\% of the evaluated malware are high-risk, and to reduce the average error in the presence of a constant maliciousness level, the samples should be estimated at $c> 0.5$, as Figure~\ref{constant} confirms.
  
   \begin{figure}
  \centering
  \includegraphics[scale=0.25]{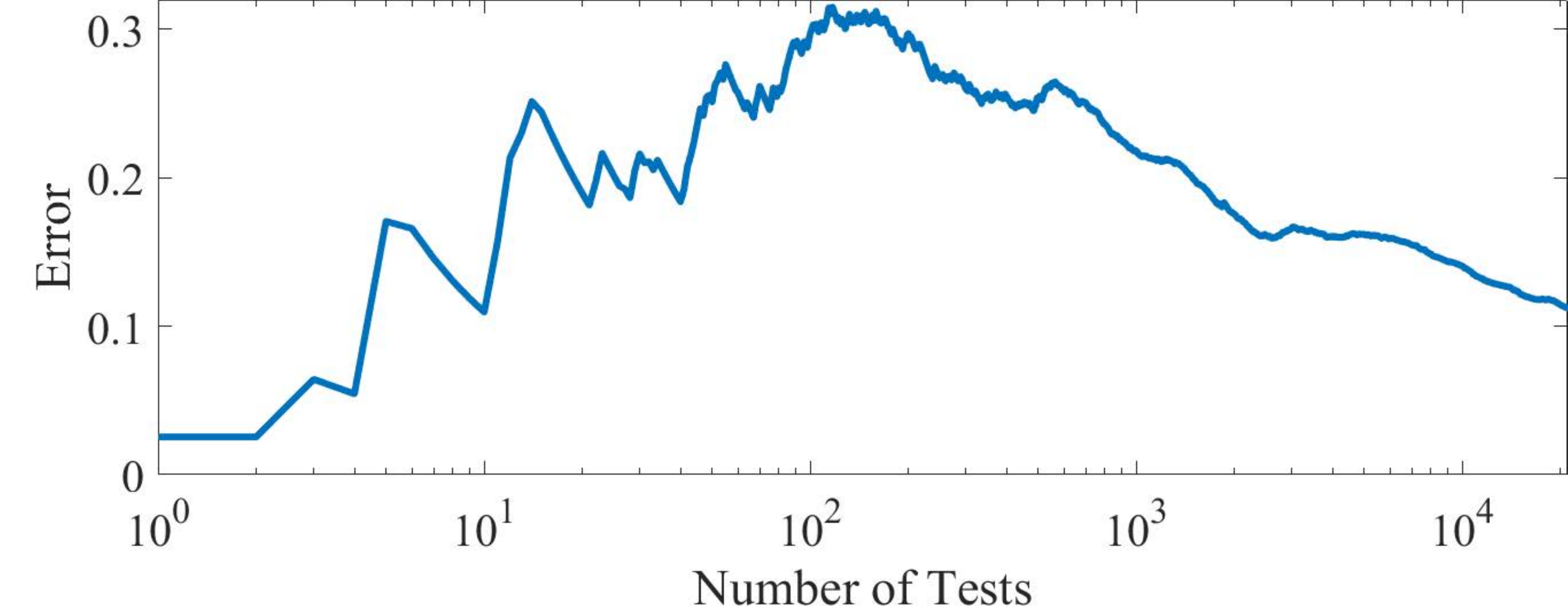}
 \caption{\small MCAE of deep learning based approach..}
  \label{cnne}
  \end{figure}
  \begin{figure}
  \centering
  \includegraphics[scale=0.37]{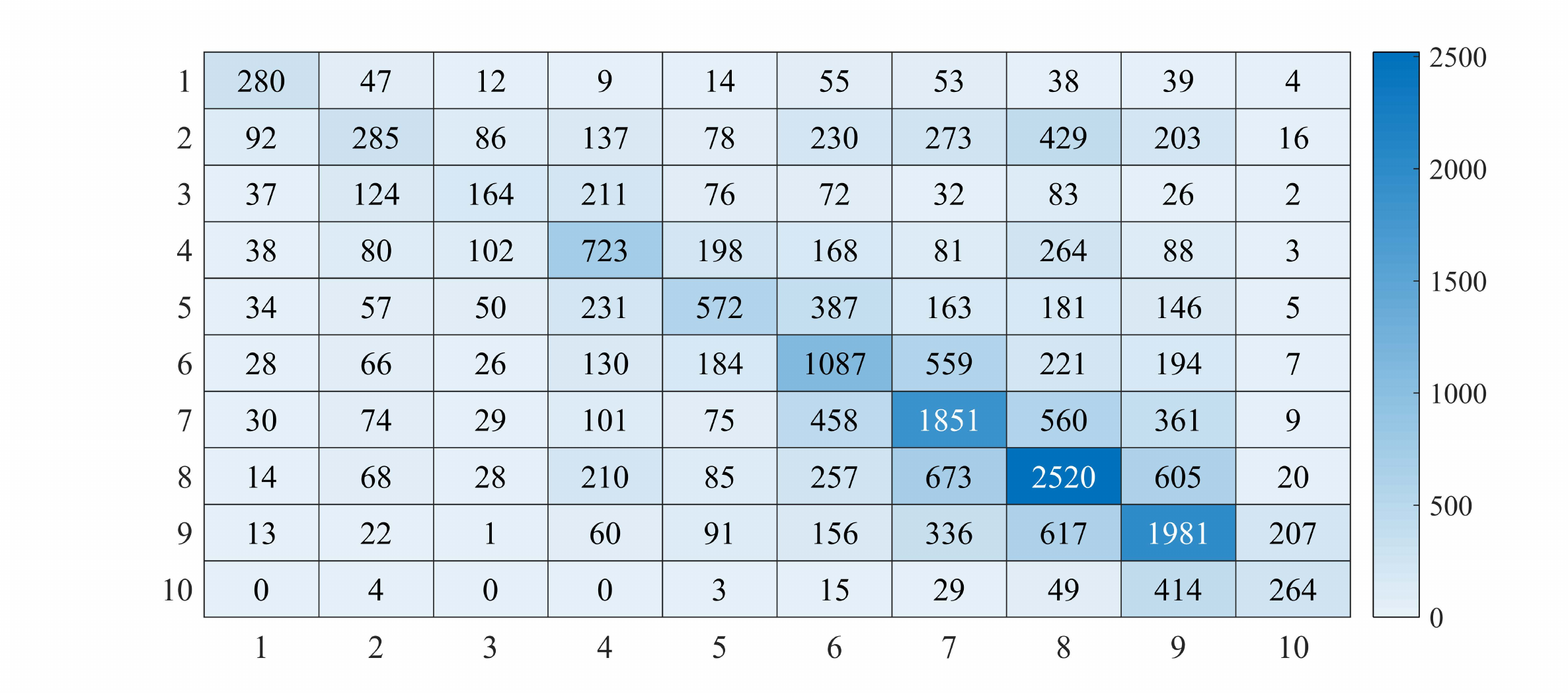}
 \caption{\small Confusion Matrix of deep learning based approach.}
  \label{cnnm}
  \end{figure} 
The results of the last experiment are provided in Figures~\ref{cnne} and \ref{cnnm}. As Figure~\ref{cnne} shows, in the first 100 tests, the average error has an upward trend, followed by a downward trend. The downtrend eventually converges to an average error of 11.1\%, which is better than all previous methods. Confusion Matrix analysis in the deep learning method also indicates the high accuracy of this method compared to other methods. As Figure~\ref{cnnm} shows, the weight of region $R_7$  is higher than other methods, which means that this method has the highest total accuracy of 45.1\%. All methods in this section have been compared using different test metrics in Table~\ref{finaltable}, except for error rate which is equal to 1 - accuracy. 

\begin{table*}[!htb]
  \centering
  \caption{\small Accuracy, Precision, Recall, F1-Score, MCAE, Conservativeness Ratio, and Total Accuracy values for evaluated dataset (Acc= accuracy; Pre= precision; Rec= recall; F1-S= F1-Score).}
\label{finaltable}
  \scriptsize{
\setlength\tabcolsep{0.5pt} 
  \begin{tabular}{|c|c|c|c|c|c|c|c|c|c|c|c|c|c|c|c|c|c|c|c|c|c|}
  \multicolumn{2}{c}{}&\multicolumn{20}{c}{\cellcolor{blue!20}\textbf{Evaluated Approach}}\\\hline
\rowcolor{yellow!50}
   \multicolumn{2}{|c|}{}&\multicolumn{4}{c|}{\textbf{Clustering}}&\multicolumn{4}{c|}{\textbf{Similarity-based}}&\multicolumn{4}{c|}{\textbf{Probabilistic}}&\multicolumn{4}{c|}{\textbf{Lattice-based}}&\multicolumn{4}{c|}{\textbf{\textbf{Deep Learning}}}\\\hline
   \rowcolor{pink!80}
    &\textbf{Maliciousness level}&\textbf{Acc}&\textbf{Pre}&\textbf{Rec}&\textbf{F1-S}&\textbf{Acc}&\textbf{Pre}&\textbf{Rec}&\textbf{F1-S}&\textbf{Acc}&\textbf{Pre}&\textbf{Rec}&\textbf{F1-S}&\textbf{Acc}&\textbf{Pre}&\textbf{Rec}&\textbf{F1-S}&\textbf{Acc}&\textbf{Pre}&\textbf{Rec}&\textbf{F1-S}\\
 \hline
\cellcolor[HTML]{FAEC0D}&\cellcolor[HTML]{E8E8AB}$0.0 \leq \xi' < 0.1$&0.518&0.046&\textbf{0.044}&\textbf{0.044}&0.855&0.04&0.006&0.01&\textbf{0.974}&-&0&0&\textbf{0.974}&-&0&0&\textbf{0.974}&\textbf{0.495}&0.013&0.026\\\cline{2-22}
\cellcolor[HTML]{FAEC0D}&\cellcolor[HTML]{E8E8AB}$0.1 \leq \xi' < 0.2$&0.864&0.047&0.007&0.012&\textbf{0.974}&-&0&0&\textbf{0.974}&-&0&0&\textbf{0.974}&-&0&0&0.903&\textbf{0.345}&\textbf{0.015}&\textbf{0.028}\\\cline{2-22}
\cellcolor[HTML]{FAEC0D}&\cellcolor[HTML]{E8E8AB}$0.2 \leq \xi' < 0.3$&0.871&0.056&0.006&0.011&\textbf{0.954}&0&0&0&\textbf{0.954}&-&0&0&\textbf{0.954}&-&0&0&\textbf{0.954}&\textbf{0.33}&\textbf{0.008}&\textbf{0.016}\\\cline{2-22}
\cellcolor[HTML]{FAEC0D}&\cellcolor[HTML]{E8E8AB}$0.3 \leq \xi' < 0.4$&0.858&0.077&0.007&0.012&\textbf{0.921}&-&0&0&\textbf{0.921}&-&0&0&\textbf{0.921}&-&0&0&0.902&\textbf{0.4}&\textbf{0.037}&\textbf{0.068}\\\cline{2-22}
\cellcolor[HTML]{FAEC0D}&\cellcolor[HTML]{E8E8AB}$0.4 \leq \xi' < 0.5$&0.81&0.038&0.005&0.009&\textbf{0.91}&0&0&0&\textbf{0.91}&-&0&0&\textbf{0.91}&-&0&0&0.904&\textbf{0.416}&\textbf{0.029}&\textbf{0.055}\\\cline{2-22}
\cellcolor[HTML]{FAEC0D}&\cellcolor[HTML]{E8E8AB}$0.5 \leq \xi' < 0.6$&0.839&0.112&0.006&0.012&0.874&\textbf{0.75}&0&0&0.874&0.435&0.004&0.008&\textbf{0.875}&-&0&0&0.851&0.377&\textbf{0.059}&\textbf{0.102}\\\cline{2-22}
\cellcolor[HTML]{FAEC0D}&\cellcolor[HTML]{E8E8AB}$0.6 \leq \xi' < 0.7$&0.824&0.238&0.006&0.011&0.752&0.153&0.024&0.042&\textbf{0.835}&-&0&0&\textbf{0.835}&-&0&0&0.819&\textbf{0.457}&\textbf{0.105}&\textbf{0.170}\\\cline{2-22}
\cellcolor[HTML]{FAEC0D}\multirow{-7}{*}{\begin{sideways}{\textbf{Estimated Label}}\end{sideways}}&\cellcolor[HTML]{E8E8AB}$0.7 \leq \xi' < 0.8$&0.758&0.464&0.016&0.03&0.754&0.317&0.006&0.012&0.263&0.246&\textbf{0.912}&\textbf{0.388}&0.76&-&0&0&\textbf{0.796}&\textbf{0.508}&0.147&0.228\\\cline{2-22}
\cellcolor[HTML]{FAEC0D}&\cellcolor[HTML]{E8E8AB}$0.8 \leq \xi' < 0.9$&0.841&0.533&0.009&0.018&0.362&0.176&\textbf{0.36}&\textbf{0.236}&\textbf{0.847}&\textbf{0.742}&0.013&0.026&0.84&-&0&0&0.834&0.488&0.11&0.18\\\cline{2-22}
\cellcolor[HTML]{FAEC0D}&\cellcolor[HTML]{E8E8AB}$0.9 \leq \xi' \leq 1.0$&\textbf{0.964}&0.4&0&0&\textbf{0.964}&\textbf{1}&0&0&\textbf{0.964}&-&0&0&0.063&0.036&\textbf{1}&\textbf{0.07}&\textbf{0.964}&0.492&0.013&0.025\\\hline
\rowcolor{green!50}
&Per class metric score &1&0&1&1&5&2&1&1&\textbf{8}&1&1&1&7&0&1&1&4&\textbf{7}&\textbf{6}&\textbf{6}\\\hline

\rowcolor{gray!20}
 \multicolumn{2}{|c|}{\textbf{MCAE}}&\multicolumn{4}{c|}{41.59\%}&\multicolumn{4}{c|}{26.36\%}&\multicolumn{4}{c|}{17.6\%}&\multicolumn{4}{c|}{36.93\%}&\multicolumn{4}{c|}{\textbf{11.1\%}}\\
\rowcolor{gray!20}
 \multicolumn{2}{|c|}{\textbf{Conservativeness Ratio}}&\multicolumn{4}{c|}{0.203}&\multicolumn{4}{c|}{3.3572}&\multicolumn{4}{c|}{3.012}&\multicolumn{4}{c|}{$\infty$}&\multicolumn{4}{c|}{1.2814}\\
 \rowcolor{gray!20}
 \multicolumn{2}{|c|}{\textbf{Total Accuracy}}&\multicolumn{4}{c|}{7.31\%}&\multicolumn{4}{c|}{15.79\%}&\multicolumn{4}{c|}{25.51\%}&\multicolumn{4}{c|}{3.63\%}&\multicolumn{4}{c|}{\textbf{45.1\%}}\\\hline
  \end{tabular}}
  \end{table*}  
 
 The bold values in Table~\ref{finaltable} represent the best results for each class, which are assigned to the estimated maliciousness levels $\xi'$'s, and "-" represents the undefined test metric. For example, in the similarity-based method $R_1, R_3 = 0$, which makes the precision at $0.1 \leq \xi' <0.2$ equal to $(0+0)/0$. As discussed earlier, the presence of zero column $C_i$ in the Confusion Matrix of an algorithm indicates that this algorithm is unable to generate output $\xi'_i$, and this appears "-" in the precision column corresponding to the $i^{th}$ estimated label.
 
 In Table~\ref{finaltable}, a \textit{per class metric score} is assigned to each approach, which shows the number of best results obtained for each metric test in different classes. For example, the accuracy of the probabilistic approach with score 8 has the best value, except for $\xi' \in [0.5,0.6)\cup [0.7,0.8)$, which is in the best rank compared to all other approaches. This score corresponds to the number of bold values for each column. Note that the total accuracy of this method is 25.51\%, which is at the second rank. The results show that except for the accuracy per class metric, deep learning is in the best position for all test metrics.  
\section{Discussions}\label{discuss}
In the previous section, several methods for malware detection were analyzed, and the proposed method was compared with similar ones. This section discusses other ideas.

In the previous section, the proposed solution was compared with the similarity-based approach, presented in \cite{similarity}. In this comparison, the samples were first converted into an image and then the method was applied to the obtained images. This approach is for detecting malware using static features, while the proposed method is based on dynamic ones. For this reason, instead of calculating the images using the proposed method, a vector corresponding to the extracted features can be used, and then the similarity-based method can be applied. In this case, the features call times are removed and the resulting vector contains the maliciousness level with the features shown in eq.~\ref{staticform}.
\begin{equation}\label{staticform}
Xi =[\xi, \phi_1 :\tau_1, \cdots, \phi_n :\tau_n] \rightarrow [\xi, \phi_1, \cdots, \phi_n ]
\end{equation}
 After this conversion, the similarity-based method can be applied to the calculated vector. Unfortunately, our implementation shows that this method suffers more errors than the image-based method, and therefore was not described in detail in the previous section.
 
The next point is about the clustering image shown in Figure~\ref{8clusters}. As mentioned earlier, this image gives us a lot of information about malware. For example, the black lines in this image indicate that there are features in the dataset that do not affect any of the clusters. Black columns, on the other hand, indicate that there are times when no feature is called. By removing these two black areas, the dimensions of the images can be reduced. Our implementation shows that this increases CNN 4 times faster. Unfortunately, this reduced the accuracy of the deep learning method by 4\%, which is why it was not covered in detail in the previous section.

The next point is to analyze the results of the lattice-based method. As stated in the previous section, the outputs of this method are always $c = 0.9828$. The reader may be asked: \textit{what this number is and why labels are always estimated with this number?}
 This number is the largest label in our trainset. To prove this, suppose our dataset has $n_ {max}$ samples with the largest maliciousness level $c$. The image form of the first instance can be represented as a binary $Z^1$ matrix. Using eq.~\ref{lam} and eq.~\ref{latt}, we can deduce the result in eq.~\ref{loutput}, which shows that a large number of zero elements in $Z^1$ increases the probability of convergence to $c$.
 \begin{equation}\label{loutput}
 Z^1_{i,j}=0 \rightarrow LAM_{i,j}=c
 \end{equation}
  Now consider the second train instance $Z^2$ with label $c$. Among the situations in which $Z^1$ contains 1, $Z^2$ may be zero. Similar to the previous case, $Z^2_{i',j'}=0$ makes $LAM_{i',j'}=c$. This continues until eq.~\ref{finalloutput} is established.
   \begin{equation}\label{finalloutput}
 \exists 1 \leq i \leq n_\phi ,1 \leq j \leq \tau ,1 \leq k \leq n_{max} \quad s.t \quad Z^k_{i,j}=0
 \end{equation}
  Note that the binary matrices are sparse with a large number of zeros. For example, the size of $Z^1$ corresponding to the maliciousness level 0.9828 in our dataset is $482 \times 60 = 28920$, while it has only 17 non-zero elements, meaning that at least 99.94\% of $LAM$'s elements are $c$. Because $c-1$ is a negative number, the remaining 17 elements will probably converge to the second largest number in the dataset, 0.9815. This analysis shows that at best, the evaluated samples are compared with a maximum of 18 maliciousness levels. Although this observation reduces the performance of the lattice-based method in detecting malware in this paper, it also has some interesting applications, which are discussed in~\cite{lattice}.
  \begin{figure}
  \centering
  \includegraphics[scale=0.22]{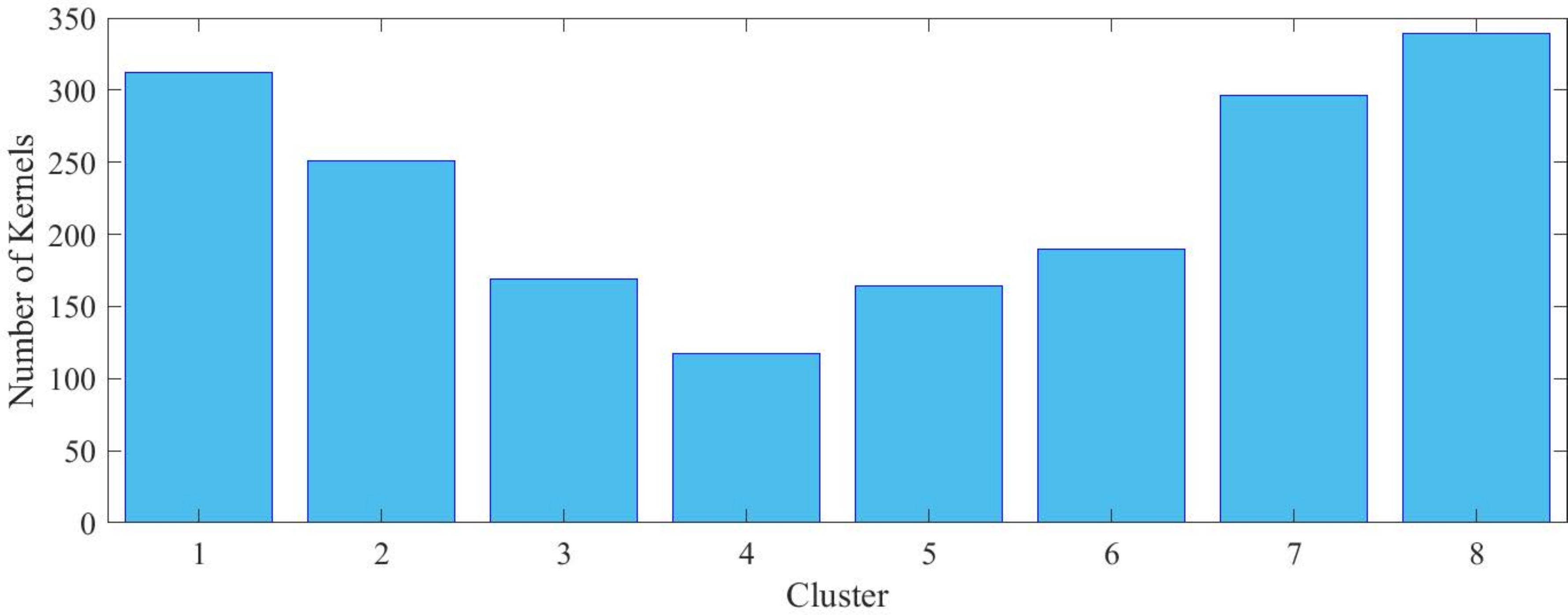}
 \caption{\small Number of kernels in different clusters.}
  \label{kernels}
  \end{figure} 
Again, see the clustering image corresponding to the 8 clusters in Figure~\ref{8clusters}. In this image, there are pixels that are common in all clusters. These points will probably impose additional overhead on the system. Our focus is on pixels that are in only one cluster, which can lead to the emergence of interesting ideas. These points are referred to as \textit{kernels} in lattice-based approach. Figure~\ref{kernels} shows the number of kernels in each cluster. In the kernel-based method, instead of using all trainset samples, only the samples containing kernels can be used, which increases the efficiency. Implementing this method improves the accuracy of the lattice-based approach from 3.6\% to 7.65\% while still not being competitive with other compared methods. Also, the conservativeness ratio of lattice-based method was $\infty$, while in the kernel-based approach it is 0.0178, meaning that the kernel-based approach is a loose method that treats high-risk samples such as low-risk ones. 
  
\section{Conclusion and Future Directions}\label{conc}
This paper introduced a method to detect online malware. The proposed method monitored the behavior of malware for a period of time, and extracts the dynamic called features during this period in the form of a vector. Due to the fact that we do not have much storage and processing space in IoT devices, proposed method in this paper transformed the extracted vector into a sparse binary image, and clustering, similarity-based, probabilistic, lattise-based, and deep learning approaches are introduced for the generated image dataset analysis. The results showed that the mean error of the clustering method is more than 40\%, which puts this method in the worst rank, and on the other hand, deep learning with an error of 11.1\% is in the best position. The evaluated methods were compared with 8 different metrics, among which the probabilistic method was in the best position in terms of accuracy per class metric, and deep learning with 45.1\% accuracy in the remaining 7 metrics surpassed others.

This paper provides valuable results for future research. For example, we have shown that MCAE is not the only effective parameter in the performance of methods, and if regions $R_1$ and $R_3$  in the Confusion Matrix of a target class are equal to zero, there will be a zero column in this matrix that makes precision metric calculation impossible. This result can be used to suggest a multi objective optimization learning base method.

The results of this article can be used in other areas such as label flipping in adversarial machine learning. For example, we showed that the lattice-based approach always returns the maximum maliciousness level $c$ as output, for the used dataset. In this case, changing label $c'< c$ will not change the performance of this method. In addition, the conservativeness ratio can be used to identify vulnerable methods against adversarial targets. If this rate is too high, the probability of identifying dangerous samples increases, while if the rate is less than 1, this method returns smaller maliciousness levels as output, and this can encourage adversaries to produce malware with high maliciousness level.

\Urlmuskip=0mu plus 1mu\relax
\bibliographystyle{IEEEtran}
\bibliography{main}

\end{document}